\documentclass[a4paper,11pt]{article}
    \usepackage{jcappub}  
    
    \pdfoutput=1
    
    \usepackage{lineno}
    \nolinenumbers
    \usepackage{booktabs}
    \usepackage{subcaption}
    \usepackage[dvipsnames]{xcolor}
    \hypersetup{
      colorlinks=true,
      linkcolor=blue,
      citecolor=purple,
      urlcolor=blue
    }
    
    \newcommand{\HI}{H\,\textsc{i}}
    
    \title{Foreground removal and 21 cm signal estimates: comparing different blind methods for the BINGO Telescope}
    
    \author[a,b,c]{Alessandro Marins}
    \author[a,b]{Filipe B.~Abdalla}
    \author[c,d,e]{Elcio Abdalla}
    \author[a,b]{Chang Feng}
    \author[f]{Luiza O.~Ponte}
    \author[g,h]{Giancarlo de Gasperis}
    \author[c]{Luiz H.~F.~Assis}
    \author[i]{Mathieu Remazeilles}
    \author[j]{Carlos A.~Wuensche}
    \author[k]{Luciano Barosi}
    \author[l]{Edmar C.~Gurj\~ao}
    \author[j,m]{Thyrso Villela}
    \author[n,o]{Bin Wang}
    \author[p]{Jiajun Zhang}
    \author[q]{Ricardo Landim}
    \author[j]{Vincenzo Liccardo}
    \author[j]{Camila P.~Novaes}
    \author[k]{Amilcar R.~Queiroz}
    \author[n]{Larissa Santos}
    \author[c,k]{Marcelo V.~dos Santos}
        
    \emailAdd{amarins@ustc.edu.cn}
    \emailAdd{fba@ustc.edu.cn}
    \emailAdd{eabdalla@usp.br}
    \emailAdd{changfeng@ustc.edu.cn}
    
    \affiliation[a]{Department of Astronomy, University of Science and Technology of China, Hefei 230026, China}
    \affiliation[b]{School of Astronomy and Space Science, University of Science and Technology of China, Hefei 230026, China}
    \affiliation[c]{Instituto de F\'isica, Universidade de S\~ao Paulo, C.P. 66318, CEP 05315-970, S\~ao Paulo, Brazil}
    \affiliation[d]{Universidade do Estado da Paraíba, R. Baraúnas 351, Universitário, Campina Grande - PB, 58429-500, Brazil}
    \affiliation[e]{Centro de Ciências Exatas e da Natureza (CCEN), Universidade Federal da Paraíba, CEP 58059-970, João Pessoa - PB, Brazil}
    \affiliation[f]{Instituto de Astronomia, Geofísica e Ciências Atmosféricas, Universidade de São Paulo, R. do Matão 1226, CEP 05508-090, São Paulo, Brazil}
    \affiliation[g]{Dipartimento di Fisica, Sapienza Universit\`a di Roma, Piazzale Aldo Moro 5, I-00185 Roma, Italy}
    \affiliation[h]{INFN Sezione di Roma, Piazzale Aldo Moro 5, I-00185 Roma, Italy}
    \affiliation[i]{Instituto de F\'isica de Cantabria (CSIC-UC), Avenida de los Castros s/n, 39005 Santander, Spain}
    \affiliation[j]{Instituto Nacional de Pesquisas Espaciais, Divis\~ao de Astrof\'isica, Av. dos Astronautas 1758, 12227-010, S\~ao Jos\'e dos Campos, SP, Brazil}
    \affiliation[k]{Unidade Acad\^emica de F\'isica, Universidade Federal de Campina Grande, R. Apr\'igio Veloso, 58429-900, Campina Grande, Brazil}
    \affiliation[l]{Unidade Acadêmica de Engenharia Elétrica, Universidade Federal de Campina Grande, R. Aprígio Veloso, 58429-900, Campina Grande, Brasil}
    \affiliation[m]{Instituto de F\'isica, Universidade de Bras\'ilia, Campus Universit\'ario Darcy Ribeiro, 70910-900, Bras\'ilia, DF, Brazil}
    \affiliation[n]{Center for Gravitation and Cosmology, Yangzhou University, Yangzhou 224009, China}
    \affiliation[o]{School of Aeronautics and Astronautics, Shanghai Jiao Tong University, Shanghai 200240, China}
    \affiliation[p]{Shanghai Astronomical Observatory, Chinese Academy of Sciences, Shanghai 200030, China}
    \affiliation[q]{Technische Universit\"at M\"unchen, Physik-Department T70, James-Franck-Strasse 1, 85748 Garching, Germany}

    \abstract{
    The BINGO radiotelescope will observe hydrogen distribution using Intensity Mapping (IM) to analyze the Dark Energy paradigm through Baryon Acoustic Oscillations. The target signal is contaminated by unwanted signals and instrumental noise, making accurate estimations essential for characterizing the 21 cm signal.
    
    In this study, we evaluated the performance of three blind foreground-removing algorithms—FastICA, GNILC, and GMCA—on the BINGO pipeline. Each method used different approaches to estimate foreground contributions, and we also investigated how the number of simulations for debiasing affects estimation quality.
    
    Our findings indicate that using 50 or 400 simulations yields equivalent results at this stage of analysis. All algorithms produced statistically consistent estimates of the 21 cm signal. We used FastICA for estimating and debiasing the \HI\ spectra from five years of observations, which yielded reliable results, although the first channel was affected by edge effects from the mixing matrix. The overall signal-to-noise ratio (SNR) was 204, and the chi-squared value ($\chi^2$) was 1.8.
    }
    
    
\begin{document}
    \maketitle
    \flushbottom
    
    \section{Introduction}

Although most post-reionization neutral hydrogen (\HI) was ionized during the Epoch of Reionization \citep{haermmerle2020}, a small fraction ($x_{\mathrm{H\,I}}\sim 2\%$) remained self-shielded inside dense systems with column densities
$N_{\mathrm{H\,I}} \gtrsim 2\times10^{20}\,\mathrm{cm}^{-2}$\footnote{$N_{\mathrm{HI}}=\int n_{\mathrm{H\,I}}\,\mathrm{d}s$, where $n_{\mathrm{HI}}$ is the \HI\ number density.},
such as damped Ly$\alpha$ absorbers (DLAs; e.g.\ \citep{wolfe2005,bird2014,mingfengho2021}).
Because DLAs host a substantial fraction of the \HI\ reservoir after reionization, they trace the distribution of neutral gas and, indirectly, the underlying matter field. The remaining \HI\ in these systems typically has a spin temperature above the CMB temperature \citep{kanekar2003}, enabling detection through the redshifted 21\,cm line.

Despite the faint 21\,cm emission from individual sources, \emph{intensity mapping} (IM) \citep{chang2010intensity-460} targets the collective, unresolved signal by integrating the radio emission over coarse angular pixels and narrow frequency channels. This approach offers a fast, efficient way to survey large cosmological volumes, at the cost of limited angular resolution. A key challenge for IM is that the redshifted 21\,cm signal is observed together with bright astrophysical and cosmological emissions (foregrounds) in the same frequency band.

Recovering the 21\,cm signal is therefore a statistical process that requires careful pre-processing and foreground mitigation. For single-dish surveys, the non-uniform angular response of the feed horns and their chromaticity complicate map-making and foreground cleaning \citep{Matshawule}. In addition, instrumental effects introduce correlated and uncorrelated noise components, further impacting the separation of foregrounds from the cosmological emission.

The BINGO telescope \citep{BINGO_I} aims to detect the 21\,cm signal over a redshift span $0.13 \lesssim z \lesssim 0.45$ (980--1260\,MHz) in the southern celestial hemisphere. BINGO will operate with 28 feed horns in a single-dish drift-scan IM mode \citep{BINGO_III}, targeting a homogeneous coverage of part of the southern hemisphere. The survey is designed to measure the baryon acoustic oscillation (BAO) feature \citep{chang2008_baoIM,novaes2022bingo-3b0}, whose detection critically depends on the fidelity of the recovered 21\,cm signal and on robust control of instrumental systematics.

Several BINGO papers have described the instrument and the current status of its analysis pipeline. The \emph{generalized needlet internal linear combination} (GNILC; \citep{Remazeilles:2011}) was successfully used in the Planck analysis (see, e.g., \citep{GNILCplanck2016}) and is currently the baseline method adopted in the BINGO pipeline for estimating the foreground contribution \citep{olivari2016,BINGO_IV,BINGO_V,mericia2023testing-a6b}. In this context, the present work extends the pipeline by implementing and testing two additional blind foreground removal approaches. Our goal is to evaluate their performance, and to identify trade-offs between accuracy and computational cost that are relevant for future applications to real data.

All algorithms considered here are blind in the sense that they do not assume an explicit physical model for the foreground emission. They exploit different statistical properties of the data: GMCA \citep{bobin2007} relies on morphological diversity and sparsity, FastICA \citep{maino2002} assumes statistical independence between non-Gaussian components (operating in pixel space), and GNILC estimates the effective mixing dimension by combining harmonic and spatial localization. At this stage, we include only uncorrelated thermal noise to maintain control over known effects, leaving frequency-dependent contributions and gain calibration (e.g.\ $1/f$ noise) for future work. We also assume a Gaussian main beam that is identical across frequency.

This paper is organized as follows. We describe the astrophysical emission components and instrumental contaminants in the data set in Section~\ref{Sec: 21cm and FG components}. The observation model is presented in Section~\ref{Sec: sky model}. The foreground-removal methods are detailed in Section~\ref{Sec: FGremoval}, and the debiasing procedure is described in Section~\ref{Sec: debias}. We introduce the statistical diagnostics in Section~\ref{Sec: statistic tools} and present the performance analysis in Section~\ref{Sec: Algorithm Estimation}. We then apply FastICA in a map-making approach in Section~\ref{Sec: HIDE}. Finally, we discuss and conclude this work in Section~\ref{Sec: Conclusions}.
    
    \section{21 cm and Foreground Components}
    \label{Sec: 21cm and FG components}
    \label{Section: 2} 
\begin{figure}
    \centering
    \includegraphics[width=\textwidth]{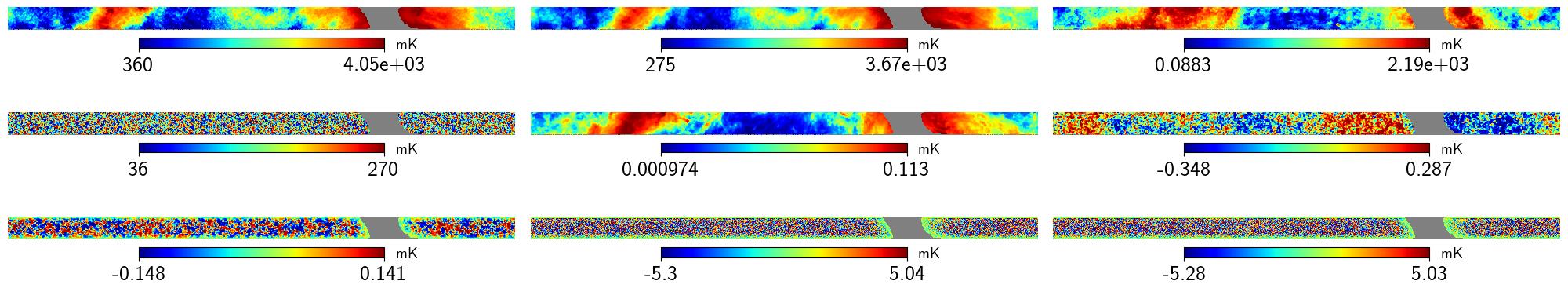}
    \caption{Intensity maps in the BINGO sky region. The maps are in antenna temperature and represent: observed (top left), synchrotron (top middle), free-free (top right), radio point sources (middle left), AME (center), CMB (middle right), HI (bottom left), thermal noise for 1-year mapping at 70 K system temperature (bottom center), and HI + thermal noise (bottom right). The gray region represents the masked part of the Galaxy.}
    \label{fig: Foregrounds}
\end{figure}

We provided detailed information about our data set in Appendix~\ref{Section: AppendixA}, which is the same set of simulations described in \citep{BINGO_V}. In Fig.~\ref{fig: Foregrounds}, we display all components used to perform the foreground-removal analysis and estimate the 21\,cm signal. Our observational signals comprise the 21\,cm signal, foregrounds, and instrumental noise, and all maps are masked with a binary mask to remove the unobserved sky.

We simulated the $\delta T_{\mathrm{H\,I}}$ fields with the \texttt{FLASK} code\footnote{\href{http://www.astro.iag.usp.br/~flask/}{FLASK webpage}}, creating 400 log-normal maps for different tomographic redshift bins. The code uses a log-normal distribution to describe the temperature fields, which is a better approximation than a Gaussian one, primarily because the log-normal distribution prevents the field from taking unrealistic negative values. \texttt{FLASK} uses the $\mathrm{H\,I}$--$\mathrm{H\,I}$ angular power spectrum generated with the \texttt{UCLCl} code \citep{mcleod2017}. As shown in \cite{BINGO_VII}, \texttt{UCLCl} generates output closer to the theoretical calculation (see Eq.~\ref{eqn: deltaT}), with a deviation of less than 1\%.

\begin{figure}
    \centering
    \includegraphics[width=\textwidth]{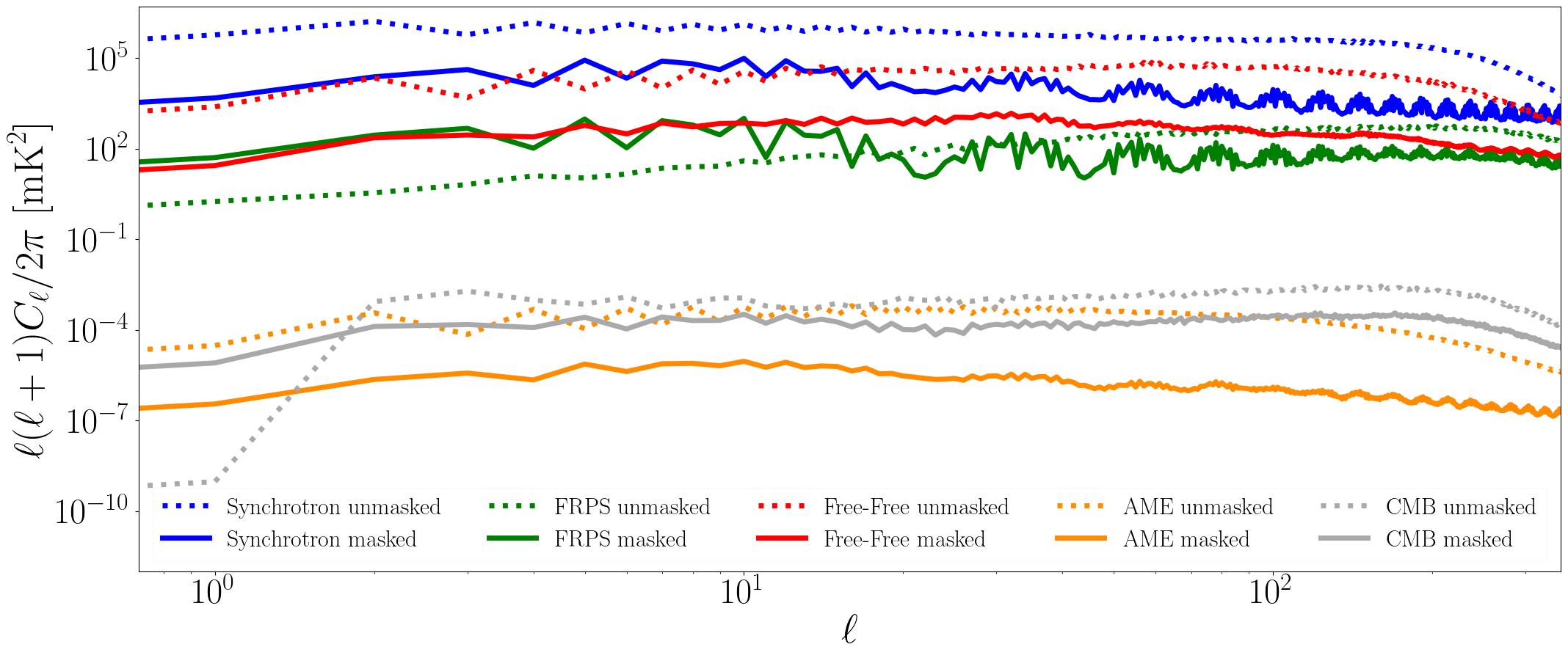}
    \caption{Angular power spectrum masked (solid line) and unmasked (dashed line) within the BINGO coverage region for all foreground components assumed in this work: CMB (gray), AME (orange), free-free (red), FRPS (green), and synchrotron (blue).}
    \label{fig: FGcls}
\end{figure}

For foregrounds, we assumed five components: synchrotron, free-free, anomalous microwave emission (AME), radio point sources (FRPS), and the CMB. Each foreground-component map was generated by the Planck Sky Model (PSM) code \citep{PSM}. Figure~\ref{fig: FGcls} displays the foreground angular power spectra for full and partial sky.

To analyze the BINGO pipeline's consistency and performance, we focused solely on thermal instrumental noise. This was modeled over one year of observations using the BINGO optical design in its Phase~1 configuration, as described in \cite{BINGO_II}. With a system temperature of $70\,\mathrm{K}$, 28 feed horns, a bandwidth of $10\,\mathrm{MHz}$, a sky-coverage fraction of $\sim 13\%$, and an ideal case of duty cycle of 100\%, each pixel is mapped for about $2.44\,\mathrm{hours/year}$, resulting in a pixel-noise level of $1.15\times0.334\,\mathrm{mK}$ in an inhomogeneous coverage.

It is crucial to note that one year of observation yields a low signal-to-noise ratio. BINGO is expected to operate for five years, yielding greater sensitivity than assumed in the first part of our analysis. In the final part of this work, we present results for five years of observation produced by the map-making process (see Section~\ref{Sec: HIDE}).

    \section{Modeling signals from the sky}
    \label{Sec: sky model}
    Drift-scan IM surveys cannot inherently distinguish between different signal sources in the sky, nor can they specifically isolate \HI\ emission. However, in radio astronomy, emissions from different parts of the sky and at different frequencies generally exhibit no correlation, as these emissions originate from the summation of independent systems lacking time or phase correlations \citep{liu2020}. Furthermore, foregrounds have smooth spectral features that can be used to estimate them and then remove them from the observation. The resulting signal is composed of the 21\,cm emission plus instrumental noise---since their near-Gaussian behavior is not modeled by blind algorithms---along with leakage terms arising from algorithmic performance \citep{marins2024}.

Sky signals can be modeled by considering contributions from the components outlined in Section~\ref{Sec: 21cm and FG components} for a specific sky position $p$ within a frequency channel $\nu_i$. The observed signal is expressed as a linear combination of contributions from independent sources.

Assuming $N_{\mathrm{s}}$ physical sources are modeled by blind algorithms and an additional unmodeled term $n$ that includes $x_{\mathrm{H\,I}}$ and $n_{\mathrm{Inst}}$ (the 21\,cm signal and instrumental noise), we can express the observation as
\begin{align}
x(\nu_i, p) = \sum_{j} y_j(\nu_i, p) + n(\nu_i, p).
\end{align}
In this expression, $y_j$ represents the $j$-th foreground component and can be factorized as $y_j(\nu_i, p) = a_{ij}\, s_j(p)$, where $a_{ij}$ describes the spectral law and $s_j(p)$ indicates the spatial response.

In a matricial representation, where each row is a \texttt{HEALPix} map \citep{HEALpix},
\begin{equation}
\mathbf{X} = \mathbf{A}\mathbf{S} + \mathbf{N},
\end{equation}
where $\mathbf{S}$ is the \emph{spatial response} matrix and $\mathbf{A}$ is the \emph{mixing matrix}. Thus, $\mathbf{A}$, $\mathbf{S}$, and $\mathbf{X}$ (and $\mathbf{N}$) have dimensions $N_{\mathrm{ch}}\times n_{\mathrm{s}}$, $n_{\mathrm{s}}\times N_{\mathrm{pix}}$, and $N_{\mathrm{ch}}\times N_{\mathrm{pix}}$, respectively. Here, $N_{\mathrm{pix}}$ is the number of pixels. While $N_{\mathrm{s}}$ represents the actual (unknown) number of foreground components, $n_{\mathrm{s}}$ is the number of templates used to model the total foreground emission. Since blind algorithms do not know \emph{a priori} the true number of sources, we assume $n_{\mathrm{s}} \in [1, N_{\mathrm{ch}}]$. Therefore, when an algorithm estimates $\hat{\mathbf{A}}$ and $\hat{\mathbf{S}}$ for a given $n_{\mathrm{s}}$, it does not necessarily recover specific astrophysical components, but rather provides an effective description of the foreground emissions.

Estimating $\mathbf{A}$ (and $\mathbf{S}$) is a linear inversion problem. Due to unmodeled signals and the fact that the mixing-matrix dimension is rarely equal to the number of channels, $\mathbf{A}$ is generally not invertible. Blind algorithms typically do not define \emph{a priori} the mixing-matrix dimension.\footnote{GNILC uses the \emph{Akaike Information Criterion} \citep{AIC1974} to estimate the dimension of the mixing matrix in both harmonic and spatial domains.} The method used to estimate this matrix characterizes each algorithm. A general solution involves finding a filter $\mathbf{W}$ that, when applied to the observation matrix $\mathbf{X}$, yields the estimated foreground matrix. This filter effectively downweights $\mathbf{N}$ while preserving the foreground signals (see Appendix~\ref{Section: AppendixB} for details).

We also define the residual matrix $\mathbf{R}$ as the difference between the observation matrix and the estimated foreground matrix:
\begin{align}
\mathbf{R} &= (\mathbf{I}-\mathbf{W}_{\mathrm{FG}})\mathbf{X},
\label{eq: algorithm_residual}
\end{align}
where $\mathbf{W}_{\mathrm{FG}} \doteq \hat{\mathbf{A}}\mathbf{W}$ denotes the foreground-reconstruction filter.

    \section{Foreground Removal methods}
    \label{Sec: FGremoval}
    To begin with, to treat the sky images, we must account for the undesirable foreground contribution to the total observed signal. We use three blind methods: the algorithms do not assume any prior knowledge about the emission sources. Their general description and assumptions are given in Appendix~\ref{Section: AppendixC}. Here, we summarize the main ideas underlying their operation.

\subsection{Fast Independent Component Analysis}
\emph{Independent Component Analysis} (ICA) is an algorithm applied to astronomical observations to model foregrounds using the hypothesis that astrophysical sources are statistically (and mutually) independent. Signals from different sources are statistically independent, meaning that they do not contain information about one another. In mathematical terms, when two signals are statistically independent, their joint probability distribution is the product of their marginal distributions.

ICA-based algorithms seek a linearly transformed matrix $\hat{\mathbf{S}}=\mathbf{W}\mathbf{X}$, where each row is a transformed vector and all components are mutually independent \citep{hyvarinen1999}. The ICA algorithm used in this paper is FastICA, a fast fixed-point algorithm that uses negentropy as a measure of non-Gaussianity. We set the non-quadratic function $g(\cdot)=\log\cosh(\cdot)$, with a maximum number of 20 iterations and a tolerance of 0.01, using the FastICA implementation from the \texttt{scikit-learn} package\footnote{\url{https://scikit-learn.org/stable/modules/generated/sklearn.decomposition.FastICA.html}} to estimate the mixing matrix and thus to estimate $\mathbf{W}$ (see Appendix \ref{Section: AppendixC} for more details).

It is worth noting that FastICA yields a unique solution to the inverse problem, with each column of $\mathbf{A}$ being unique up to the signal.

\subsection{Generalized Morphological Component Analysis}
The Generalized Morphological Component Analysis (GMCA) method is based on \emph{sparsity} and \emph{morphological diversity}. A signal is sparse if it can be represented by a small number of elements from a collection of parametrized waveforms. An important example is the one provided by wavelets on the sphere (for instance, starlets), which are well localized in both harmonic and pixel domains. The method allows more than one class of waveforms to represent the observation, while sparsity reduces the effective number of components and supports morphological diversity.

In this work, we used only starlets\footnote{Also known as the \emph{Isotropic Undecimated Wavelet Transform} on the sphere.} \citep{starckIUWT2007}, which are widely used in astronomical image analysis because they capture features from astrophysical objects that are, in most cases, approximately isotropic in many applications. The use of starlet here is motivated by its application and previous tests, working efficiently in a sparse description of galactic diffuse emissions \citep{carucci2020}.

\subsection{Generalized Needlet Internal Linear Combination\label{sec:GNILC0}}
The effective ratio between the 21\,cm signal and the total observed signal depends on the pixels (i.e.\ sky direction). This variability implies that the number of (non-physical) templates required to accurately describe foreground contributions varies across regions, especially when comparing the Galactic plane neighborhood to regions at high galactic latitudes (b $\gtrsim 30^{\circ}$).

Thermal noise is Gaussian, and for low redshift, the 21\,cm signal is expected to be nearly Gaussian as well. However, foreground signals typically do not follow a Gaussian distribution, and this difference influences their relative contributions across various frequencies. Consequently, it is essential to have a localized understanding of the impact of foreground signals in both spatial and spectral domains. By moving some of the analysis to harmonic space, we can utilize localized filtering techniques such as needlets \citep{NeedletsCMB2008}.

The \emph{Generalized Internal Linear Combination} estimates a multidimensional ILC filter in both pixel and bandwidth domains by projecting the set of maps onto needlet space, modeling the observation as a set of multidimensional components (templates) rather than a single template.

We used the needlets as described in \citep{NILC2013}, through a set of cosine filters with \emph{bandcenters}\footnote{The bandcenters determine the peak, minimum, and maximum multipoles in the needlet filters described in \citep{NILC2013}.} 0, 128, and 383, each corresponding to a finite multipole range (\emph{needlet bands}). Each needlet band decomposes the observation into different angular scales.

\subsection{General observations}
We initially used all maps in all channels with \texttt{HEALPix} resolution of $\mathrm{NSIDE}=512$, convolved with a Gaussian main beam of $\mathrm{FWHM}=40$ arcmin. The convolved maps are then degraded to $\mathrm{NSIDE}=256$. This is an idealized assumption since not all feed-horn beams have an exact Gaussian shape. Although the central feed horn in the focal plane has a nearly Gaussian main beam, the farther a feed horn is from the center of the focal plane, the less Gaussian its main beam becomes. Furthermore, the construction of the reflectors also alters the feed-horn beams. The BINGO reflectors will be built modularly, connecting panels of size $\sim 1\times 2~\mathrm{m}^2$. This construction affects the beam shape, mainly through the more distant lobes. The BINGO optical design is described in \citep{BINGO_III}; the impact of realistic reflector surfaces and assembly details will be described in a future work.

    \section{Noise debias process}
    \label{Sec: debias}
    After subtracting the estimated foreground contribution from the multi-frequency maps, we obtain a residual data set that contains the 21\,cm signal plus instrumental noise (Eq. \ref{eq: algorithm_residual}). We generated 400 independent realizations of simulated 21\,cm and noise maps.

For a given method and for the $i$th realization, we estimate the 21\,cm angular power spectrum as
\begin{equation}
\widehat{C}_{\ell,i}^{\mathrm{H\,I}}
=
\frac{C_{\ell,i}^{\mathrm{R}}}{S_{\ell,i}}
-
\left\langle C_{\ell,j}^{\mathrm{N}}\right\rangle_{j\neq i},
\label{eq: Cl_HI_est}
\end{equation}
where $C_{\ell,i}^{\mathrm{R}}$ is the power spectrum of the residual map (i.e.\ the algorithm residual) for realization $i$, and $\langle C_{\ell,j}^{\mathrm{N}}\rangle_{j\neq i}$ is the average noise power spectrum computed over all realizations except $i$. The factor $S_{\ell,i}$ accounts for the multiplicative suppression of power induced by the foreground-removal step (i.e.\ loss of modes), and is estimated from simulations as
\begin{equation}
S_{\ell,i}
=
\left\langle
\frac{C_{\ell,j}^{\mathrm{R}}}{C_{\ell,j}^{\mathrm{H\,I+N}}}
\right\rangle_{j\neq i},
\label{eq: suppression_factor}
\end{equation}
where $C_{\ell,j}^{\mathrm{H\,I+N}}$ is the input (true) spectrum of the corresponding simulated \HI\ + noise maps.

    \section{Jackknife variance}
    \label{Sec: statistic tools}
    We used a resampling method to estimate the variance of each angular power spectrum recovered after foreground removal and noise debiasing. We tested both \emph{jackknife} and \emph{bootstrap} methods \citep{shao2012_jackknife,konishi2014_jackknife} and we did not identify any relevant distinction between them. We therefore adopted the jackknife because it is computationally faster. We applied a $\chi^2$-per-multipole test to assess the statistical quality of the results.

\subsection{Jackknife variance estimator}
\label{sec:jack}
The Jackknife method is a resampling strategy that works by sequentially deleting one observation at a time in a given data set and then recomputing the desired statistics. In our case, we want to estimate the variance of the angular power spectrum for each multipole and channel. We thus define a vector containing the angular power spectrum with fixed multipole $\ell$ and channel $\nu$ and all, except the specific $i$-th, $N$ different realizations.
\begin{equation}
\mathbf{C}_{\ell}^{(-i)} \doteq \left(C_{\ell,0},\ldots,C_{\ell,i-1},C_{\ell,i+1},\ldots,C_{\ell,n_r-1}\right)^{\mathrm{T}}.
\end{equation}
Thus, we can build $n_r$ new samples from the original one. Let $G(\cdot)$ denote a generic estimator computed from a sample. The corresponding jackknife estimate is
\begin{equation}
\hat{\theta}^{(-i)}_{(\ell)} = G\!\left(\mathbf{C}^{(-i)}_{\ell}\right).
\end{equation}
In this work, we take $G$ to be the mean estimator.

It is convenient to collect the estimator values over the multipole range into a vector,
\begin{equation}
\hat{\boldsymbol{\Theta}}^{(-i)} =
\left(\hat{\theta}^{(-i)}_{(\ell_{\mathrm{min}})},\ldots,\hat{\theta}^{(-i)}_{(\ell_{\mathrm{max}})}\right)^{\mathrm{T}}.
\end{equation}

The jackknife estimate of the covariance matrix is then
\begin{equation}
\mathrm{Cov}_{\mathrm{JK}}
=
\frac{n_r-1}{n_r}
\sum_{i=0}^{n_r-1}
\left(\hat{\boldsymbol{\Theta}}^{(-i)}-\left\langle\boldsymbol{\Theta}\right\rangle\right)
\left(\hat{\boldsymbol{\Theta}}^{(-i)}-\left\langle\boldsymbol{\Theta}\right\rangle\right)^{\mathrm{T}},
\label{eq: jackknife_cov}
\end{equation}
where $\langle \boldsymbol{\Theta} \rangle$ is the mean estimator over all jackknife samples. The jackknife variance at multipole $\ell$ corresponds to the diagonal element $\sigma_{\mathrm{JK},\ell}^{2}=\left[\mathrm{Cov}_{\mathrm{JK}}\right]_{\ell\ell}$.

\subsection{The $\chi^2$ test}
To quantify our results, we use the $\chi^2$ test as a measure of the goodness-of-fit of the model to the estimated data. We also use the jackknife variance described in Section~\ref{sec:jack} (Eq.~\ref{eq: jackknife_cov}) as an estimator of the variance for each angular power spectrum, per multipole and per channel. Thus, for each algorithm (GNILC, GMCA, FastICA) and a specific realization $i$, we calculate
\begin{equation}
\chi^{2}_i(\nu,\ell)
=
\frac{\left(\widehat{C}_{\ell,i}^{\mathrm{H\,I}}(\nu)-C_{\ell,i}^{\mathrm{input}}(\nu)\right)^2}
{\sigma_{\mathrm{JK},\ell}^{2}(\nu)}.
\label{eq: xi2}
\end{equation}
From this expression, it is straightforward to compute averages over frequency ($\chi^{2}_{\mathrm{eff}}$), over multipoles ($\chi^{2}_{\ell}$), and over both ($\chi^{2}_{\mathrm{overall}}$).

    \section{Algorithm's performance}
    \label{Sec: Algorithm Estimation}
    \label{Section: 7}
\begin{figure}[tbp]
\centering
\begin{subfigure}{0.48\textwidth}
  \centering
  \includegraphics[width=\linewidth]{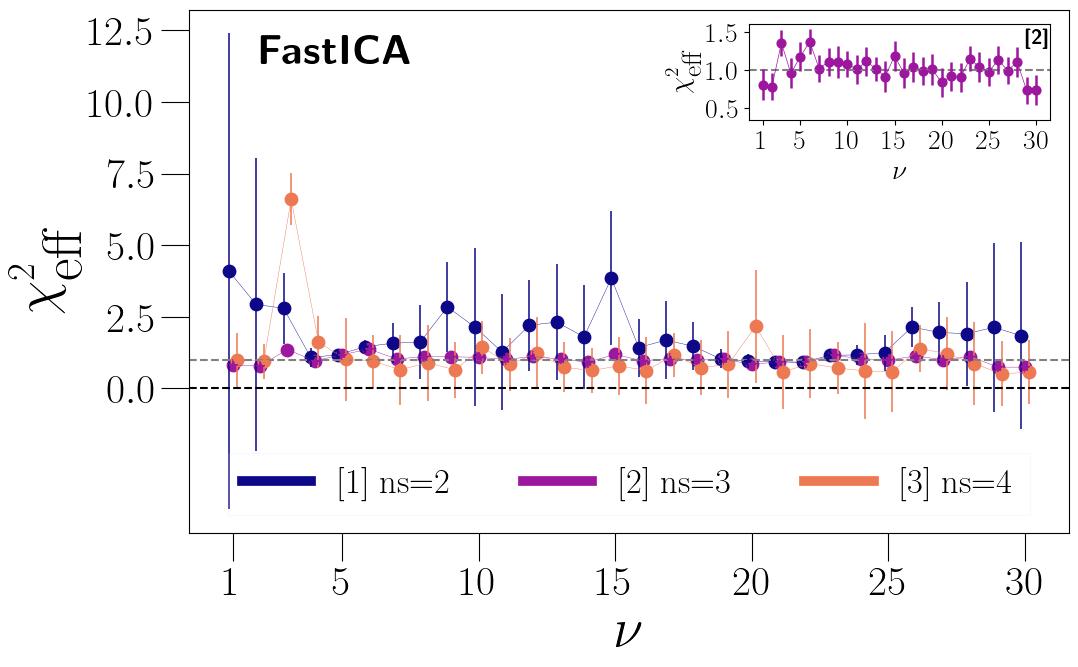}
  \label{fig: chi2_ica}
\end{subfigure}
\hfill
\begin{subfigure}{0.48\textwidth}
  \centering
  \includegraphics[width=\linewidth]{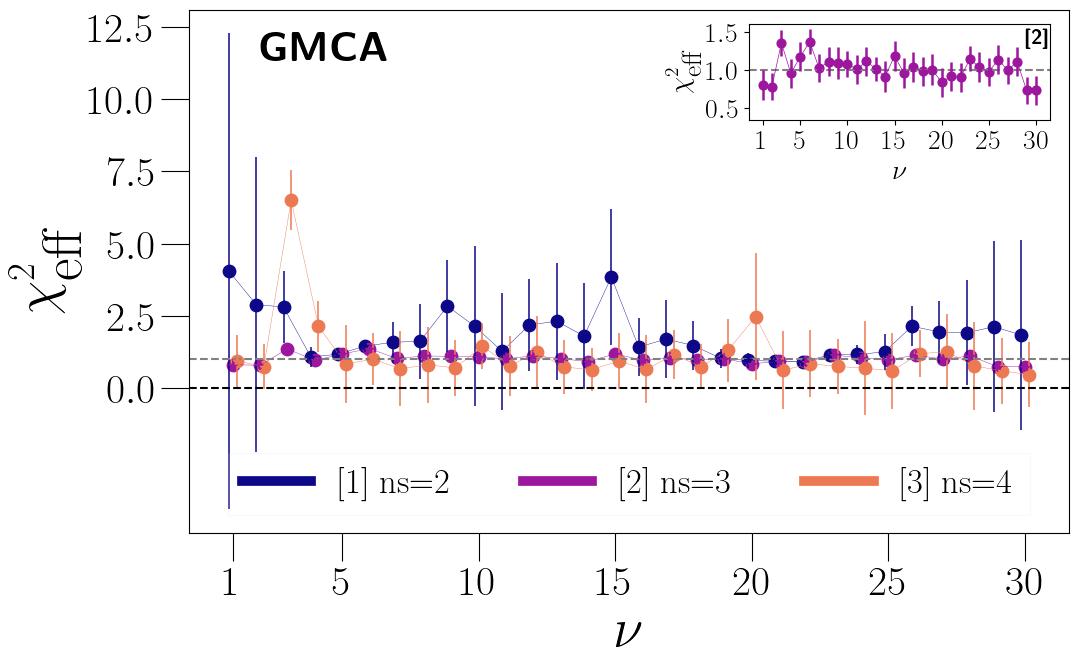}
  \label{fig: chi2_gmca}
\end{subfigure}
\caption{Effective $\chi^2$ (i.e.\ $\chi^2$ per channel) for mixing-matrix dimensions from 2 to 4, comparing \textbf{(left)} FastICA and \textbf{(right)} GMCA. In each panel, the smaller plot is a zoom-in of the main plot to highlight the behaviour of the $n_{\mathrm{s}}=3$ curve. Error bars are jackknife uncertainties from 400 realizations. The black and grey dashed lines are for reference only.}
\label{fig: chi2_eff_ns_compare}
\end{figure}
To illustrate the influence of the main foreground components on the observational map's variance, Figure \ref{fig: FGcls} presents the angular power spectrum for each foreground component in our dataset for full-sky (dashed) and constrained to the BINGO sky region (solid). By masking a significant amount of Galactic emission, the contribution of extragalactic point sources increases, which is illustrated by the dashed green curve, especially at large scales (for multipoles $\ell<20$). The decrease in amplitude is caused by two primary factors: the sample variance size ($\sim\! f_{\rm{sky}}$) and smoothing over structures in the power spectrum on scales scaling with $\Delta\ell\sim 1/\Delta\rm{DEC}$ (declination width covered in radians) \citep{tegmark1996method-eb5,tegmark1997cmb-7b9}.

\subsection{Influence of mixing matrix dimension on the GMCA and FastICA estimations}
\begin{table}
\centering
\caption{$\chi^2_{\mathrm{overall}}$ tests for the recovered spectra. Top panel: dependence on the mixing-matrix dimension $n_{\mathrm{s}}$ (400 realizations). Bottom panel: dependence on the number of realizations $n_r$ (fixed $n_{\mathrm{s}}$; see note).  Each result includes its standard deviation in parentheses.}
\label{tab: chi2overall_combined}

\medskip
\noindent\textbf{Panel A: varying $n_{\mathrm{s}}$ (fixed $n_r=400$).}
\smallskip

\begin{tabular}{lcc}
\hline\hline
$n_{\mathrm{s}}$ & FastICA & GMCA \\
\hline
2 & 1.8(4)  & 1.8(4)  \\
3 & 1.02(4) & 1.02(4) \\
4 & 1.1(2)  & 1.1(2)  \\
\hline
\end{tabular}

\medskip
\noindent\textbf{Panel B: varying $n_r$ (fixed $n_{\mathrm{s}}=3$ for FastICA/GMCA).}
\smallskip

\begin{tabular}{lccc}
\hline
$n_r$ & GMCA & FastICA & GNILC \\
\hline
400 & 1.02(4) & 1.02(4) & 1.04(4) \\
200 & 1.04(4) & 1.01(4) & 1.04(4) \\
100 & 1.03(3) & 1.01(4) & 1.03(3) \\
 50 & 1.01(3) & 0.99(3) & 1.01(3) \\
 25 & 0.97(4) & 0.95(3) & 0.97(4) \\
\hline\hline
\end{tabular}

\end{table}
\begin{figure}
    \centering
    \includegraphics[width=\columnwidth]{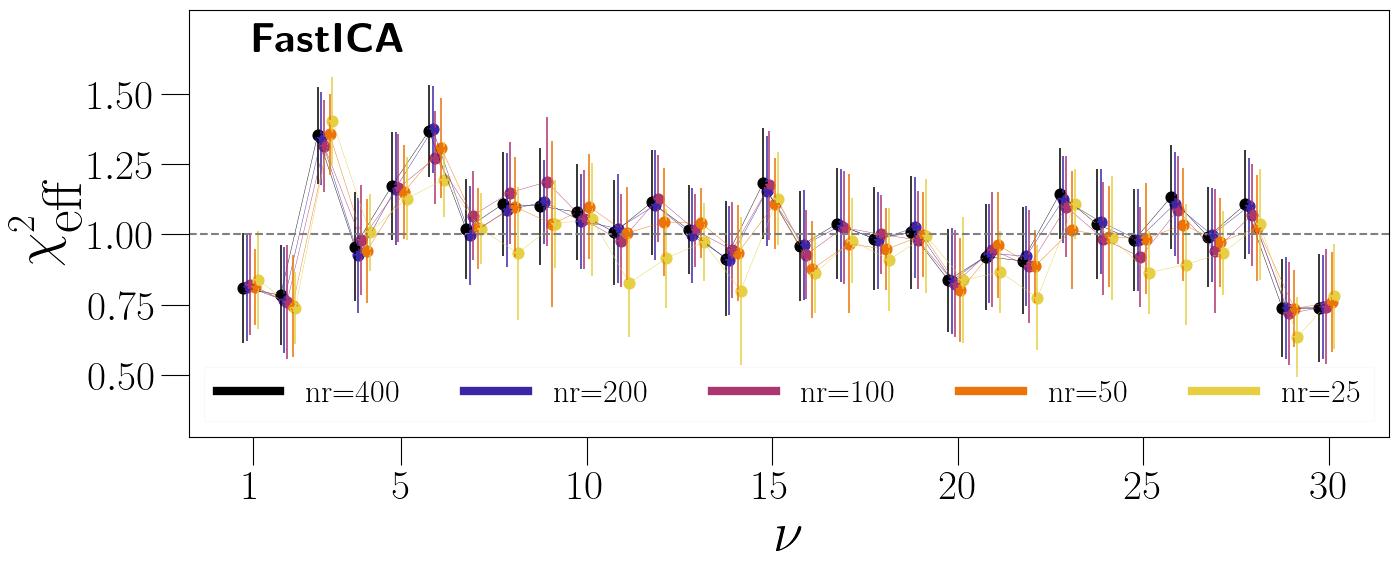}\\
    \includegraphics[width=\columnwidth]{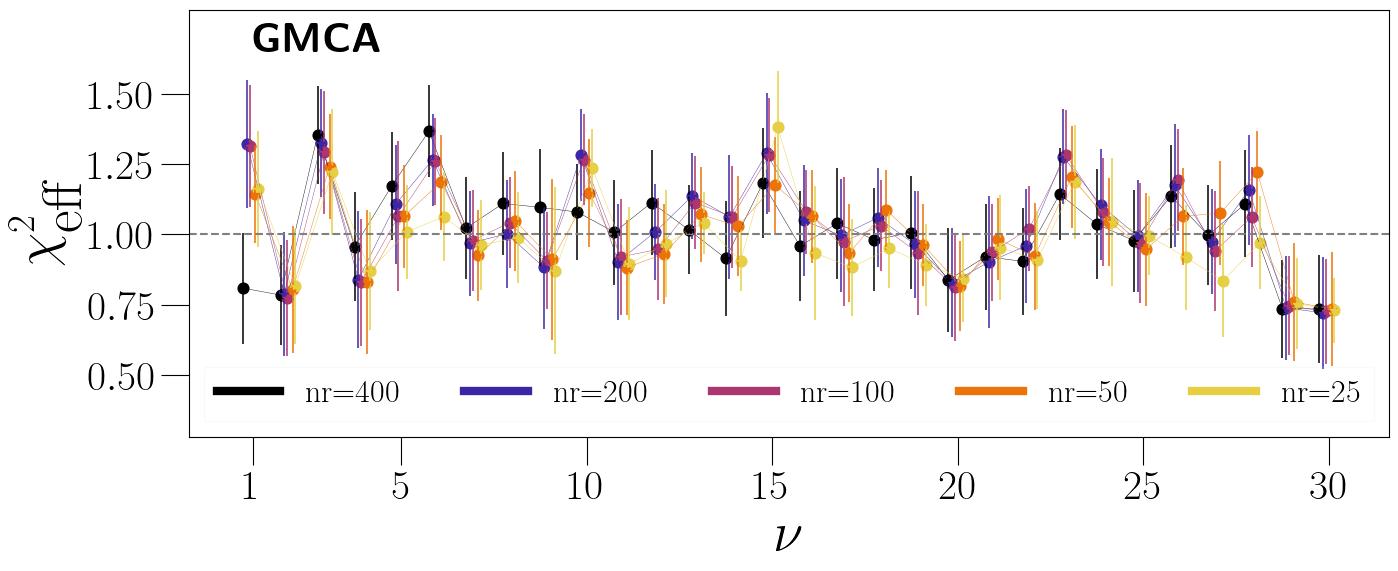}\\
    \includegraphics[width=\columnwidth]{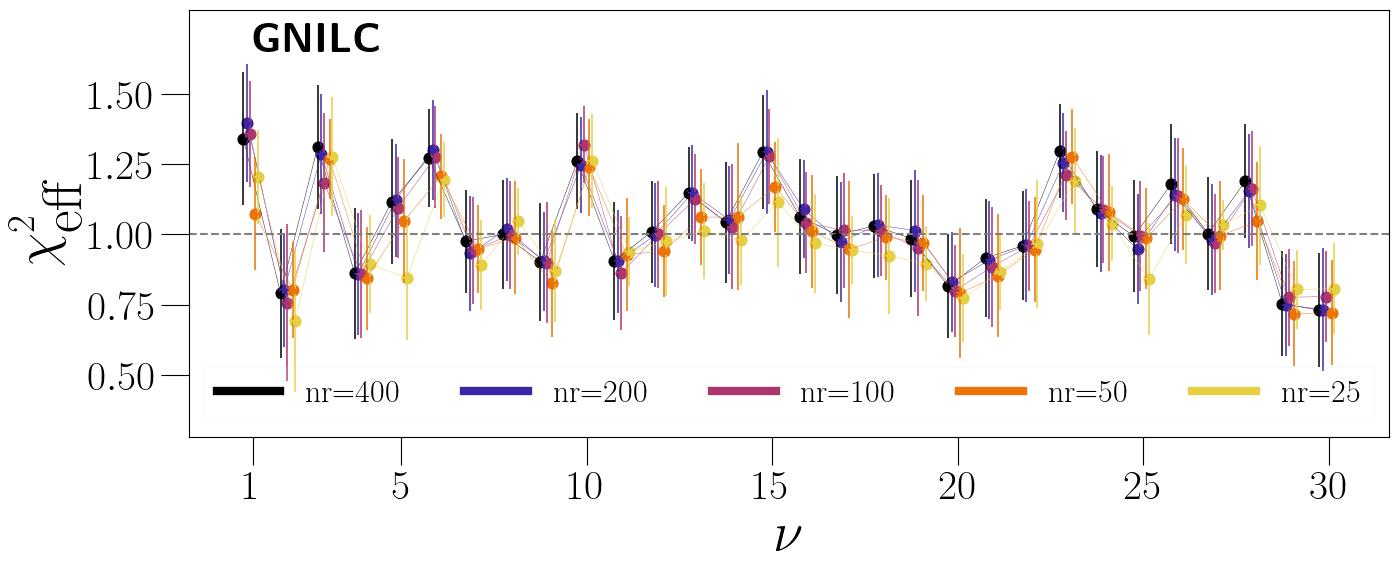}
    \caption{Effective $\chi^2$ per channel for the three algorithms. The results are for five numbers of simulations from 25 (yellow) to 400 (black).}
    \label{fig: chi2_eff_nr_CS}
\end{figure}
In contrast to the GNILC algorithm, FastICA and GMCA do not have a native way to measure the foreground contribution on the covariance of the observational maps (i.e., $N_{s}$)\footnote{Note $N_s$ is the number of foreground components of the simulated sky, as described in sec. \ref{sec:GNILC}, and $n_s$ is the number of non-physical templates assumed to represent the total foreground signals.}. 
GNILC uses the AIC (Akaike Information Criterion) \citep{AIC1974} method per needlet band and per pixel. On the other hand, for GMCA and FastICA, we assumed the number of (normalized) eigenvalues contributing to most of the total observational variance. 

The eigenvalues of the sky maps' covariance matrix are largely dominated by foregrounds, with few components accounting for most of the data variance \citep{liu2012}, since the foreground parametric description contains some parameters that are highly degenerate due to their spectral lack of features, except for point sources. However, care is required, as this is the case in which we assume no other instrumental influence. For real observational data, other systematics and man-made noise compose the data, and a higher number of components is required to remove the contaminants \citep{chang2010intensity-460, wolz2022}.

We used three configurations for FastICA and GMCA, with $n_s$ ranging from 2 to 4, and 400 realizations. The results were expressed in terms of $\chi^2_{\textrm{eff}}$ in Figure \ref{fig: chi2_eff_ns_compare}, where the variance was estimated by Jackknife, and summarized in Panel A of Table \ref{tab: chi2overall_combined}. The higher value and variance for the case with two components is a result of its higher foreground leakage into the algorithm's residual. Both algorithms led to similar results.

\subsection{Influence of the number of realizations on the estimations}
We also varied the number of simulations to estimate how much our results could be biased by the samples' size\footnote{FastICA and GMCA set as $n_s=3$.}.

Figure \ref{fig: chi2_eff_nr_CS} presents the results for five different numbers of simulations. In examining the values from FastICA and GMCA values, we observe that, for 400 realizations, both algorithms exhibit a statistically similar shape, as we can confirm from Panel B in Table \ref{tab: chi2overall_combined}.  The table also indicates that only the case with 25 realizations shows incompatibility for the FastICA algorithm. For all other simulation counts, the algorithms demonstrate statistical compatibility. Notably, GMCA and FastICA exhibit identical shapes with 400 realizations. As the number of realizations increases, both algorithms converge towards the same shape and the same overall $\chi_ {\textrm{overall}}^2$ value

\subsection{Comparison between GNILC, GMCA, and FastICA algorithms}
\begin{figure*} 
    \centering
    \includegraphics[scale=0.23]{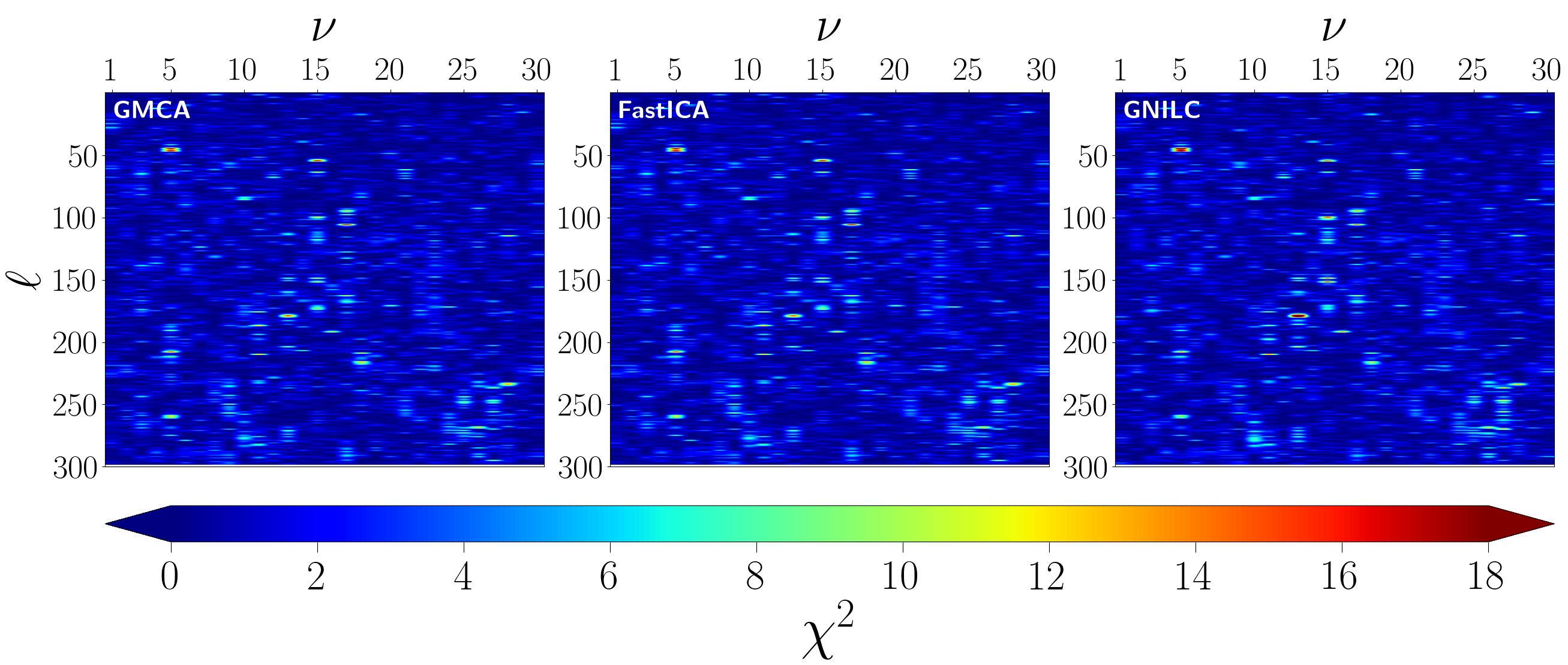}
    \caption{The plots are the $\chi^2$ (Eq. \ref{eq: xi2}) heat map of the three algorithms. The horizontal axis corresponds to the channel, and the vertical one to the estimated \HI\, angular power spectrum corresponding to a specific multipole. The redder values concerning \HI\, angular power spectrum are harder to estimate.}    
    \label{fig: heatmapXi2}
\end{figure*}
\begin{figure}
\centering
    \begin{subfigure}{0.49\textwidth}
        \centering
        \includegraphics[width=\linewidth]{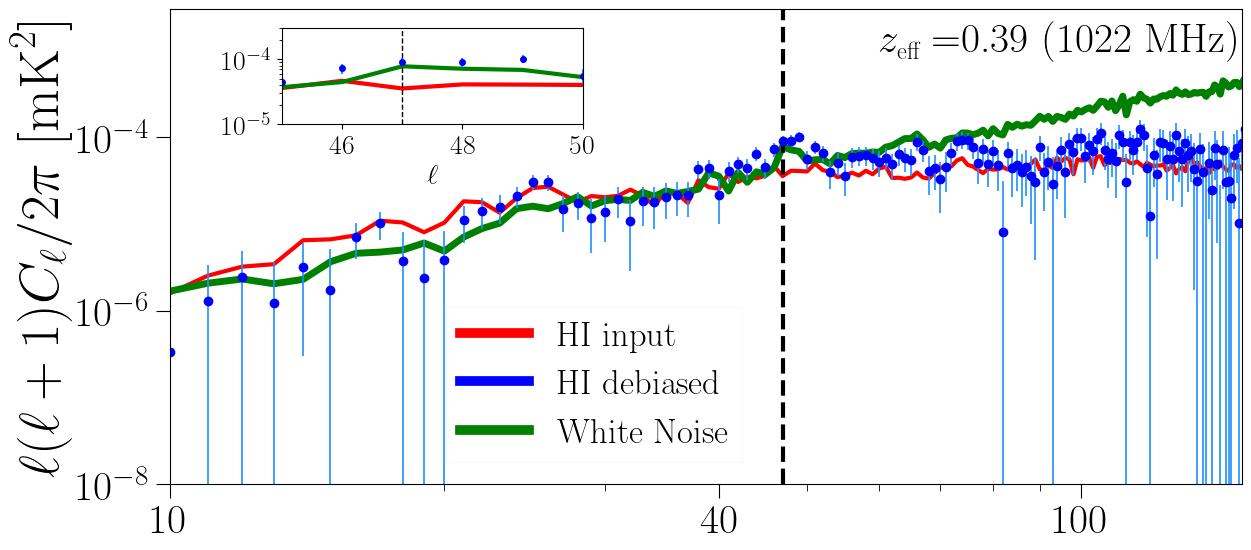}
        \caption{5th channel}
        \label{fig: prob_ch5}
    \end{subfigure}
    \hfill
    \begin{subfigure}{0.49\textwidth}
        \centering
        \includegraphics[width=\linewidth]{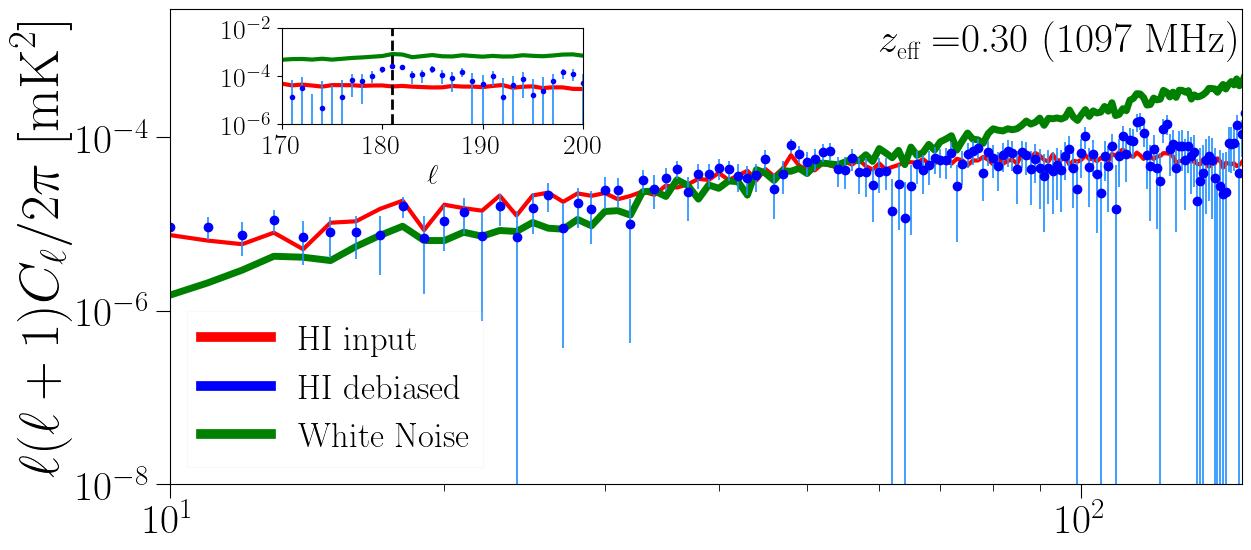}
        \caption{13th channel}
        \label{fig: prob_ch13}
    \end{subfigure}

    \caption{Angular power spectra using GMCA with $n_r = 3$ and 400 realizations for the \textbf{(left)} 5th channel and \textbf{(right)} 13th channel. The different x-axis scales in each plot better illustrate the deviation between the estimated \HI\ signal (blue) and the input (red). The green color represents the white noise angular power spectrum. The vertical black dashed line corresponds to the multipole 47 (top plots) and 181 (bottom plots).}
    \label{fig: problems}
\end{figure}

The 21 cm angular power spectrum estimated for the tenth channel is shown in Figure \ref{fig: Algorithms_HI_APS_estimation}. A $\chi^2$ heatmap for 400 simulations is in Figure \ref{fig:  heatmapXi2}. The figure is a three-dimensional representation of the reconstruction and indicates good reconstruction for all algorithms within the analyzed regime. Some regions where there are highlighted redder values are due to the low signal-to-ratio given the short observation time (one year), as around $(\nu,\ell) = (4, 47)$ and $(12, 181)$ at the top and bottom in Figure \ref{fig: problems}, respectively.

\begin{figure} 
    \centering
    \includegraphics[width=\columnwidth]{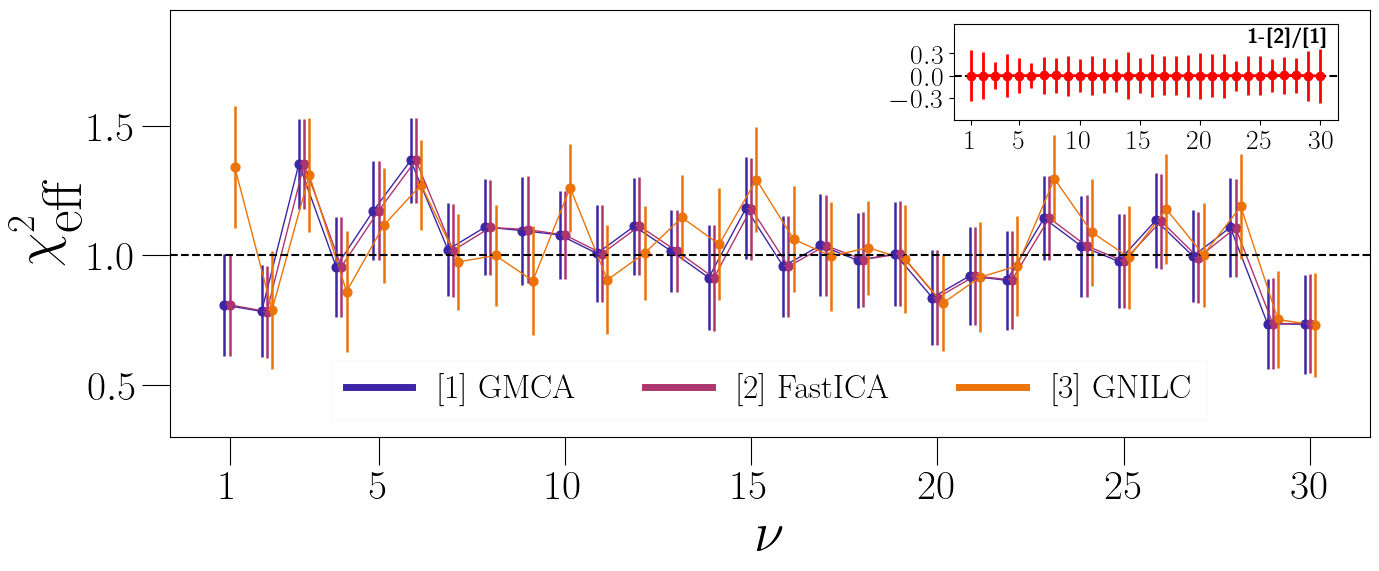}
    \caption{At the top, there is a comparison of effective $\chi^2$ of the GMCA (solid blue line), FastICA (dashed orange line), and GNILC (solid gray line). At the bottom, we represent the residual difference between the effective $\chi^2$ of FastICA and GMCA.}
    \label{fig: chi2_eff_CS_comparison}
\end{figure}

In Figure \ref{fig: chi2_eff_CS_comparison}, we can see similar results for GMCA and FastICA, as previously pointed out. In the smaller top-right plot, we show the percentage ratio between the latest algorithms, for which no statistical difference is observed. At a map level, the algorithms do not estimate the foreground contribution equally, but both carry the same statistical information. FastICA foreground estimation is slightly higher in the Galactic region, likely due to its estimation in pixel space. 

In order to verify if there is a slight preference for some of the algorithms, we calculated the difference in their AIC values from their overall chi-squared as  
\begin{table}
\centering
\caption{Pairwise differences in the Akaike information criterion, $\Delta\mathrm{AIC}$, between foreground-removal methods. Here $\Delta\mathrm{AIC}$ is equivalent to the difference in the overall $\chi^2$; positive values favour the method with the larger $\chi^2$. Uncertainties are estimated with the jackknife technique. All pairwise comparisons are consistent with zero.}
\label{tab:deltaAIC_pairwise}
\begin{tabular}{lccc}
\hline\hline
 & GMCA & FastICA & GNILC \\
\hline
GMCA    & 0       & 0.00(5)  & $-0.02(6)$ \\
FastICA & 0.00(5) & 0        & $-0.02(6)$ \\
GNILC   & 0.02(6) & 0.02(6)  & 0 \\
\hline\hline
\end{tabular}
\end{table}

\begin{eqnarray}
\Delta\textrm{AIC}_{jk} = \chi^{2}_{j} - \chi^{2}_{k}
\label{eq: Delta_AIC}
\end{eqnarray}
$j$ and $k$ are the algorithms ($j\neq k$). $\Delta\textrm{AIC}_{jk}>0$, which can be represented as "evidence in favor" of the algorithm $j$, compared to the $k$ algorithm. Table \ref{tab:deltaAIC_pairwise} presents no evidence in favor of any of the algorithms. 

In terms of computational cost, the FastICA algorithm is the most effective for processing an observational map, typically taking less than 1 minute per run on an \emph{Intel(R) Xeon(R) CPU E5-2640 v4} operating at 2.40 GHz. In comparison, the GMCA and GNILC algorithms require significantly more time, taking about 6 to 10 minutes to complete the same task.
    
    \section{Naive map-making}
    \label{Sec: HIDE}
    \begin{figure}
    \centering
    \includegraphics[width=\textwidth]{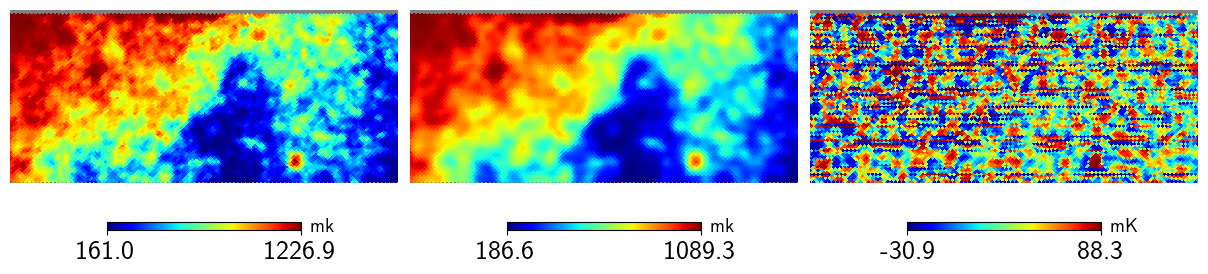}
    \caption{Left: sky emission; middle: \texttt{HIDE} output; right: difference between them, with the sky map smoothed with a Gaussian beam of 40 arcmin. The maps correspond to a mean redshift and frequency of 0.27 and 1115 MHz, respectively. The sky area corresponds to right ascension between $65^{\circ}$ and $100^{\circ}$, and declination between $-25.3^{\circ}$ and $-9.9^{\circ}$.}
    \label{fig: HIDE_SKY_diff}
\end{figure}

\begin{figure}
    \centering
    \includegraphics[width=0.8\textwidth]{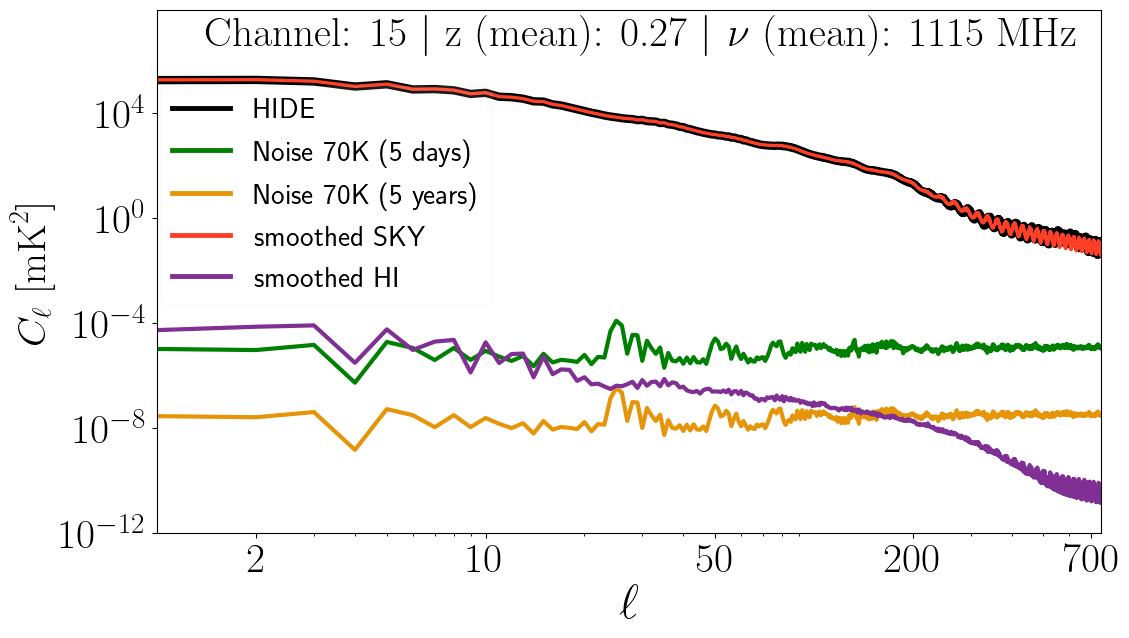}
    \caption{Angular power spectrum of the input sky (red), input \HI\ (purple), and the sky map processed by \texttt{HIDE} (black). The mask used is a combination of a mask for the telescope's region coverage and another that removes 20\% of the brightest pixels from the sky map. The \HI\ spectrum crosses the noise for five days (green) at $\ell\approx 10$ and the noise for five years at $\ell\approx 200$.}
    \label{fig: cl_HIDExINPUT}
\end{figure}

To process the observational maps in our pipeline, we have adapted the \texttt{HIDE} part of the \texttt{HIDE}\&\texttt{SEEK} software\footnote{\href{https://github.com/cosmo-ethz/hide}{https://github.com/cosmo-ethz/hide}} for BINGO specifications. \texttt{HIDE} is an efficient open-source software for simulating the observational mapping of a single-dish radio-telescope survey, producing time-ordered data (TOD). Adapting \texttt{HIDE} for BINGO has been a collaborative effort, with a dual purpose: cross-checking the current pipeline and taking advantage of the \texttt{HIDE}\&\texttt{SEEK} architecture to include new modules (beam models, RFI, noise models, etc.) and interfaces to other software and codes (\texttt{FRBlib} \citep{BINGO-IX}, \texttt{TICRA-GRASP}\footnote{\href{https://www.ticra.com/software/grasp/}{https://www.ticra.com/software/grasp/}}, \texttt{GEM} \citep{GEM}, etc.). From the \texttt{HIDE} TOD, we have built the sky-covered maps using a naive map-making associating the ordinary average of the integrated signal for each pixel.

In the \texttt{HIDE} application for BINGO, we adopted certain approximations. To more accurately reflect the actual BINGO configuration, the sky coverage used here differs slightly from previous versions. We initially analyzed the methods using the configuration assumed in earlier work \citep{BINGO_V}. Subsequently, we applied the insights gained from that analysis in the next steps. This step takes into account the initial and final positions of the feed horns on the focal plane, in a design called double-rectangular, as detailed in \citep{BINGO_III}.

The BINGO focal plane allows each feed horn to move vertically within $\pm 15$ cm to $\pm 30$ cm over five years. This movement enables distinct declination points to be targeted each year, promoting coverage homogeneity across 140 declination positions, as analyzed in \citep{BINGO_IV}.

The setup ensures smooth operation throughout the five-year span, incorporating all possible declination positions and assuming only instrumental white noise for one year of mapping at each horn location. We fixed the beam shape to be Gaussian with $\mathrm{FWHM}=40$ arcmin, corresponding to our worst-case angular resolution and matching the resolution at which we downgrade the maps.

The declination range has been adjusted to between -25.3$^{\circ}$ and -9.9$^{\circ}$ using the double-rectangular arrangement. This change has resulted in a reduction of the percentage difference in sky coverage to less than 0.04\%. The primary effect of this adjustment is related to the mask applied to the 20\% strongest emissions from the sky, which overlap with the coverage in a different region. The current sky coverage does not affect the results obtained previously.

We produced 100 sky realizations. For each realization, we generated time-ordered data (TOD) for one year of mapping, out of a total of five years of mapping, by varying the feed horn pointing and mapping. From these TODs, we created hitmaps and intensity maps for both 5 days and 5 years. The hitmaps were constructed by counting the number of times each pixel was observed. The intensity maps were derived from a straightforward map-making process, where for each pixel, we integrated the associated brightness temperature and normalized it by the corresponding hitmap value. 

Figure~\ref{fig: HIDE_SKY_diff} illustrates the results for one sky realization at a mean frequency of 1115 MHz (redshift $z\simeq 0.27$). The right panel shows the difference map between the original sky (smoothed at 40 arcmin) and the \texttt{HIDE} output, revealing striping associated with the incomplete sky coverage in the pixelization resolution and following the scanning strategy. The corresponding angular power spectra are presented in Fig.~\ref{fig: cl_HIDExINPUT}.

This step does not include gain-calibration effects or radio-frequency interference (RFI), which will be presented in a separate work. Based on the results from the previous sections, the \texttt{HIDE} maps were processed in pixel space using FastICA. After estimating the foreground emission for all simulations, we obtained \HI\ estimates and then debiased them using the same procedure described in Section~\ref{Sec: debias}.

\begin{figure}
    \centering
    \includegraphics[width=0.49\textwidth]{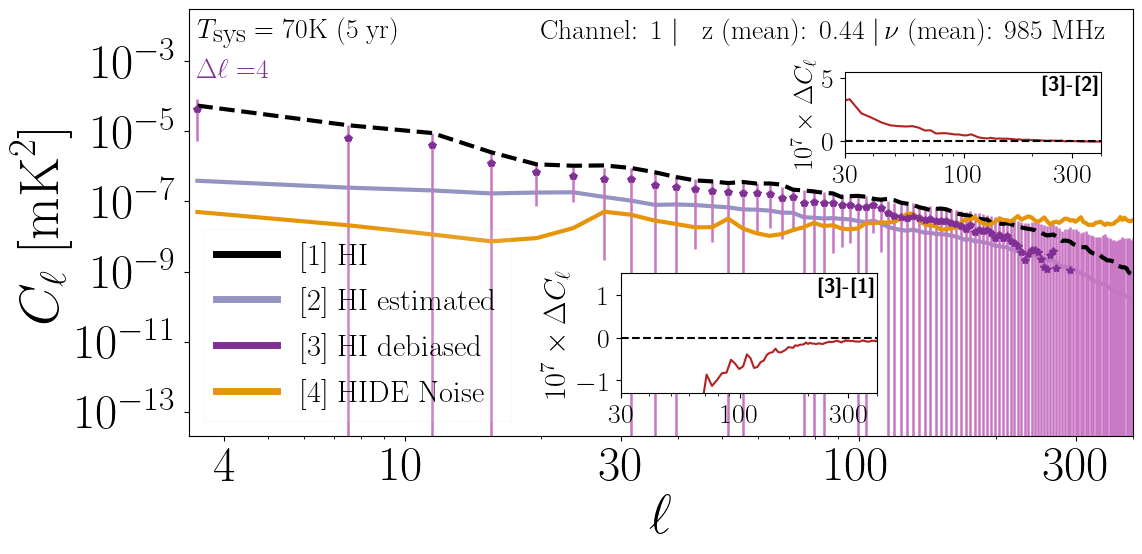}
    \includegraphics[width=0.49\textwidth]{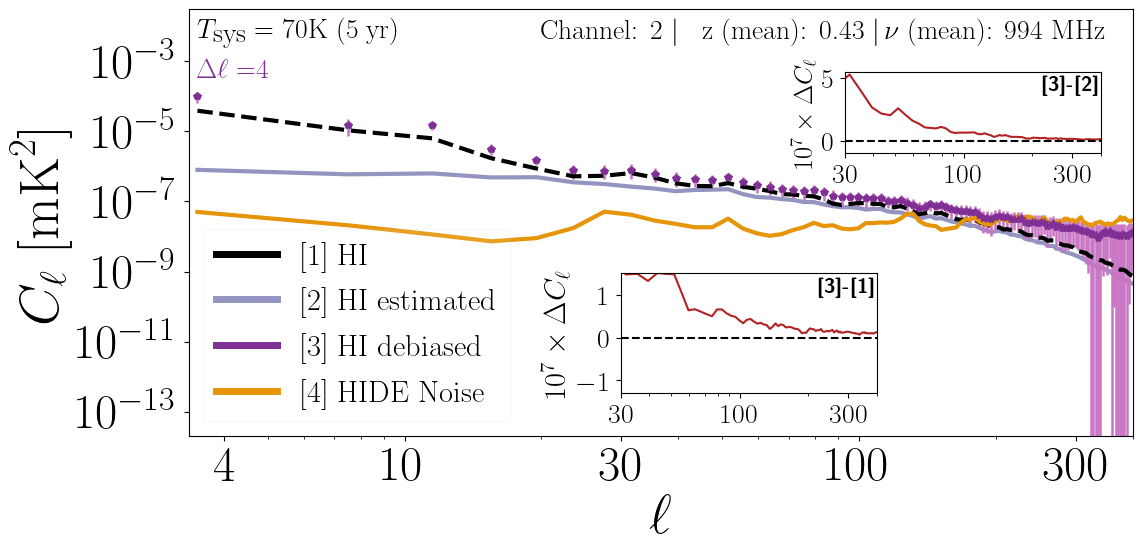}\\
    \includegraphics[width=0.49\textwidth]{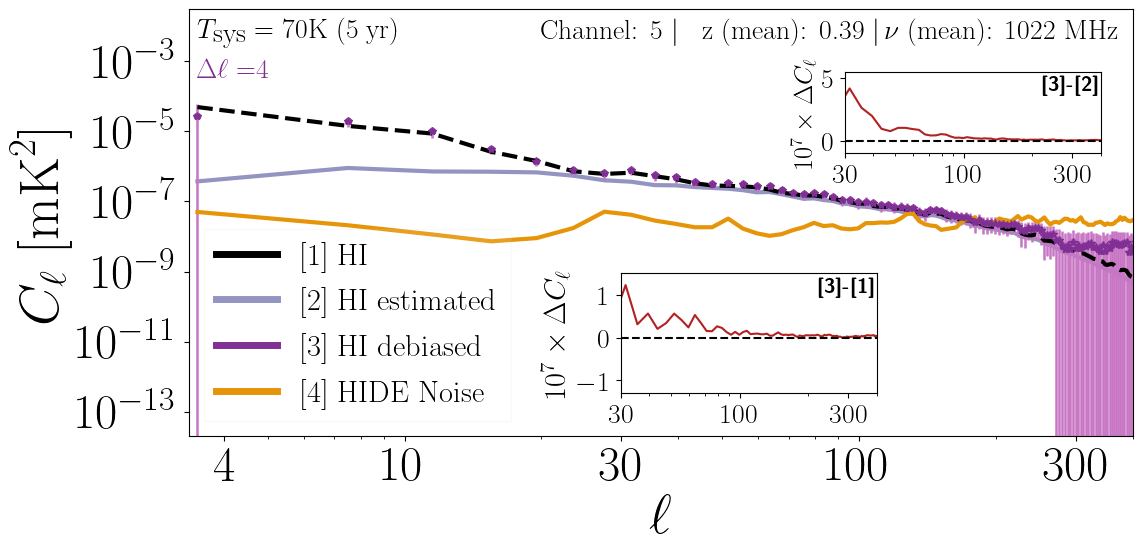}
    \includegraphics[width=0.49\textwidth]{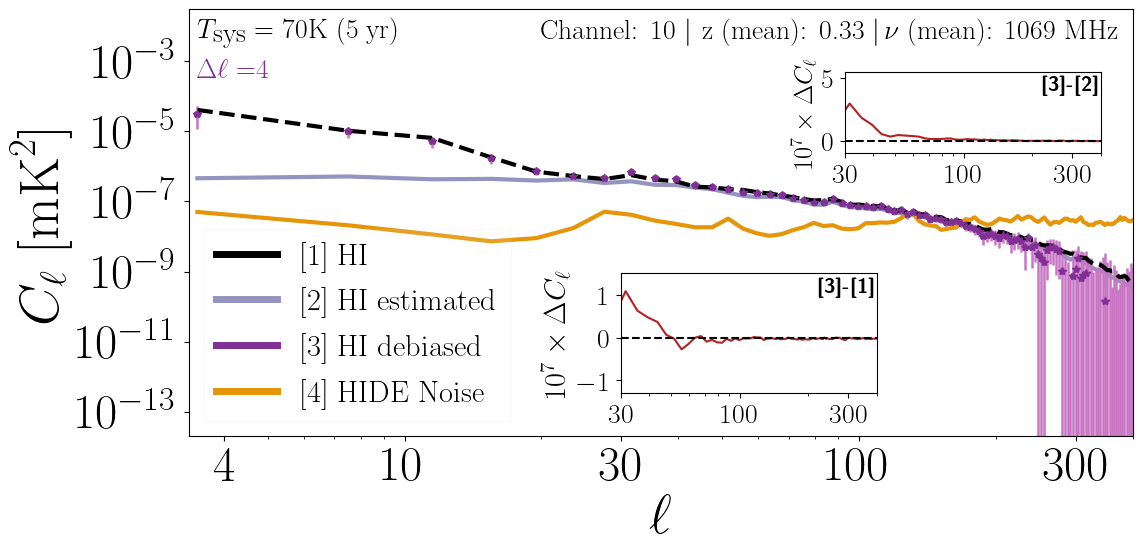}\\
    \includegraphics[width=0.49\textwidth]{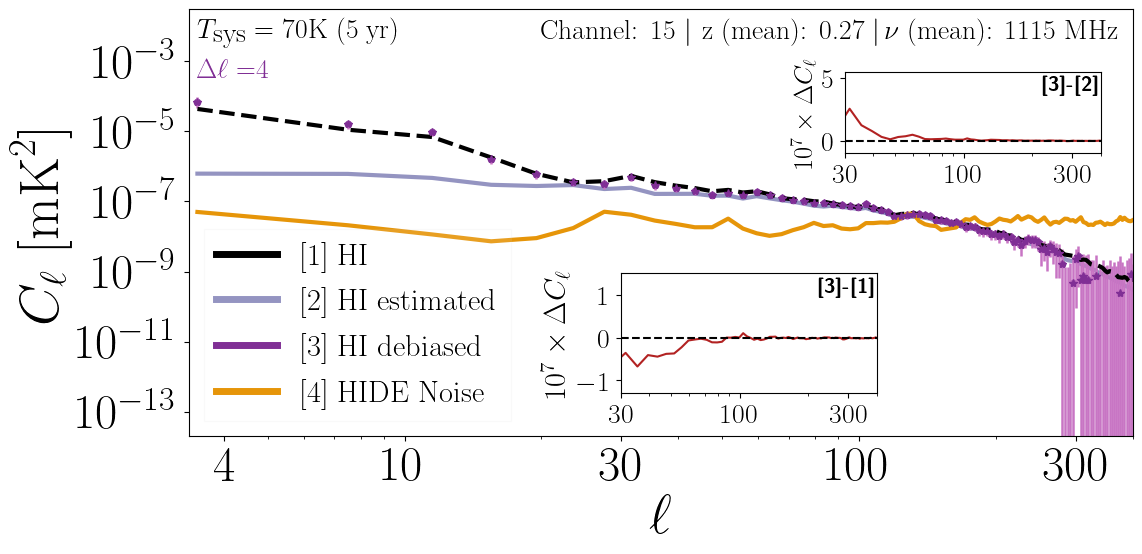}
    \includegraphics[width=0.49\textwidth]{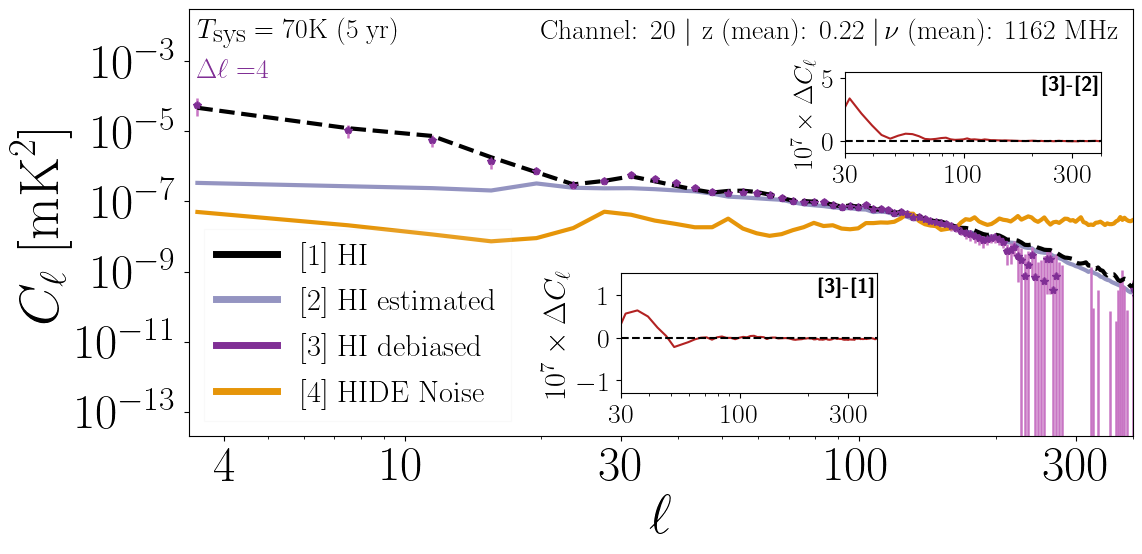}\\
    \includegraphics[width=0.49\textwidth]{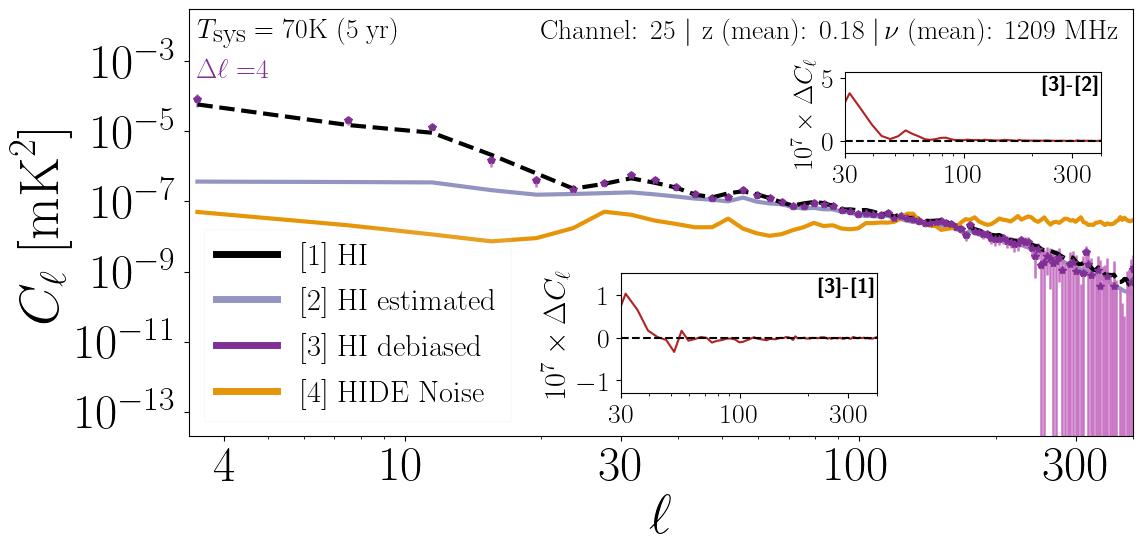}
    \includegraphics[width=0.49\textwidth]{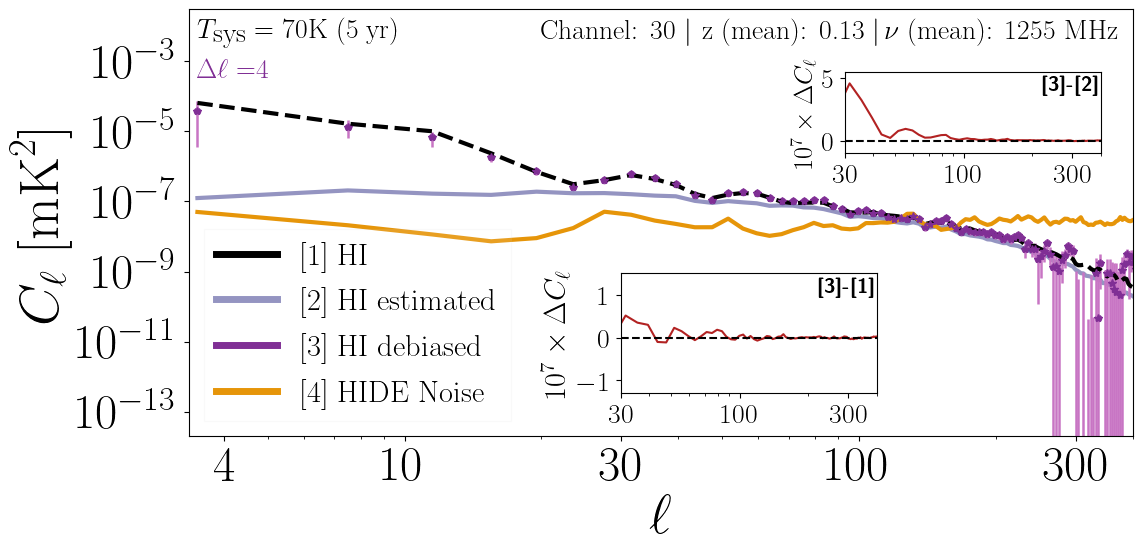}
    \caption{Angular power spectra of the input \HI\ (dashed black), \HI\ from foreground removal (slate-blue), and debiased \HI\ (purple). The debiased \HI\ is binned with multipole interval $\Delta\ell=4$. The \texttt{HIDE} noise spectra for five years at $T_{\mathrm{sys}}=70$ K (orange) cross the \HI\ curves at $\ell \sim 150$. In each panel, the top-right inset compares debiased and biased \HI, and the bottom-middle inset shows the difference between debiased and target \HI.}
    \label{fig: cl_HIDE_HI_estimated}
\end{figure}

Figure~\ref{fig: cl_HIDE_HI_estimated} presents the results for eight frequency channels for five-year observations. The spectra are binned in multipole with $\Delta \ell = 4$ (approximately $1/\Delta\mathrm{DEC}$). The \texttt{HIDE} noise (orange) corresponds to the sensitivity level for one year of operation at each horn location. As shown, the multipole at which the noise becomes dominant (low signal-to-noise ratio compared to \HI) is around $\ell \sim 150$--200.

The first channel, associated with the highest redshift, performs worst due to the mixing-matrix \emph{edge effect} and the limited foreground spectral information \citep{marins2024}. Considering scales $3 \le \ell \le 300$, the total SNR for this channel is low ($\sim 6$) as well as its $\chi^2$ ($\sim 0.2$), but this is compensated by subsequent channels, with SNR increasing from $\sim 20$ to $\sim 50$. Channels beyond the fifth (redshifts lower than 0.39) provide the best results, once they avoid residuals from the edge effect, especially between $\ell \sim 10-200$, where noise begins to dominate.  Within scales such as $\ell\sim 3-400$, their SNRs are between $30-47$ per channel in a $\chi^2\sim1-4$. Overall, we find ${\rm SNR}\simeq 204$ and $\chi^2\simeq 1.8$.

    \section{Conclusions}
    \label{Sec: Conclusions}
    \label{sec:conclusions}

We presented a controlled comparison of three blind foreground-removal methods---FastICA, GMCA, and GNILC---using simulated multifrequency maps within the current BINGO analysis pipeline, including foreground emission and thermal noise. We assessed the recovered \HI\ angular power spectra using a simulation-based debiasing procedure, jackknife error estimates, and $\chi^2$ diagnostics.

For FastICA and GMCA, we investigated the impact of the assumed mixing-matrix dimension. Adopting $n_{\mathrm{s}}=3$ foreground templates provided the best overall performance, with $\chi^2_{\mathrm{overall}}\simeq 1.02$ for both methods. Using only two templates led to increased foreground leakage and a poorer fit, consistent with the fact that a small number of dominant foreground degrees of freedom controls the covariance structure in the BINGO frequency range under our idealized instrumental assumptions.

We also quantified the number of simulations required for stable debiasing of the reconstructed \HI\ spectra. Results obtained with $n_r=50$ and $n_r=400$ realizations were statistically consistent for all algorithms, while $n_r=25$ produced an incompatible reconstruction for FastICA. This indicates that the debiasing step can be substantially accelerated without degrading accuracy in the present setup.

Across the tested multipole and frequency ranges, all three methods yielded statistically consistent \HI\ reconstructions. Pairwise differences in information criteria were consistent with zero, indicating no preferred method based solely on reconstruction accuracy.

Motivated by computational cost, we implemented FastICA in a map-making validation using \texttt{HIDE} time-ordered data tailored to BINGO. For five years of observations, we recovered the \HI\ angular power spectra with high confidence on scales where the signal is comparable to or exceeds the noise. In particular, we obtain good signal-to-noise for $\ell\simeq 3$--300 at redshifts $z\lesssim 0.39$.

    \section*{Acknowledgments}
    AM and FBA acknowledge the support from the University of Science and Technology of China. FBA also acknowledges support from the Chinese Academy of Sciences and the Br-A Talent Program.
    The BINGO project is supported by FAPESP grant 2014/07885-0; the support from CNPq is also gratefully acknowledged (E.A.). L.O.P. acknowledges the FAPESP grants 2025/16334-2 and 2023/07564-9. 
    C.A.W. thanks CNPq for grants 407446/2021-4 and 12505/2022-1, the Brazilian Ministry of Science, Technology and Innovation (MCTI) and the Brazilian Space Agency (AEB) who supported the present work under the PO 20VB.0009.
    MR acknowledges support by the Spanish Ministry of Science and Innovation (MCIN) and the Agencia Estatal de Investigación (AEI) through the project grants PID2022-139223OB-C21 and PID2022-140670NA-I00.
    
    \appendix
    
    \section{Observational components}
\label{Section: AppendixA}

\subsection{21 cm signals}
Atomic neutral hydrogen (\HI) emits a characteristic signal due to the hyperfine splitting of the $1S$ ground state. This signal has a wavelength of $21\,\mathrm{cm}$, corresponding to the frequency $\nu_{10}=1420\,\mathrm{MHz}$ in the rest frame. Despite the very low transition probability ($\sim 10^{-15}\,\mathrm{s}^{-1}$), in an astrophysical context \HI\ clouds contain a sufficient amount of \HI\ to be excited---either radiatively or collisionally---and emit detectable radiation \citep{field1958}.

Let $A_{10}$, $m_{\mathrm{H\,I}}$, $n_{\mathrm{H\,I}}$, and $\Omega_{\mathrm{H\,I}}$ be the spontaneous-emission coefficient of the 21\,cm transition, the \HI\ atom mass, the \HI\ number density, and the \HI\ density parameter, respectively. Following \citep{Furlanetto2006} and \citep{pritchard2012}, we can write the observed (average) 21\,cm brightness temperature at low redshifts, $z<2$, as
\begin{align}
T_{\mathrm{H\,I}}(z)
&=
\left(\frac{9hc^3A_{10}}{256\pi^2G k_{\mathrm{B}}\nu_{10}^2 m_{\mathrm{H\,I}}}\right)
\frac{\Omega_{\mathrm{H\,I}}(z)}{(1+z)^2}
\frac{H_{0}^2}{\left| dv_{\parallel}/d\chi \right|}
\,,
\label{eq: Tb_mean}
\end{align}
where $G$, $c$, and $h$ are the gravitational constant, the speed of light, and Planck's constant, respectively; $H_0$ is the Hubble constant; and $\left|dv_{\parallel}/d\chi\right|$ is the gradient of the peculiar velocity along the line of sight. In comoving coordinates, $\chi=a^{-1}s$, with scale factor $a=a(z)$.

Different mechanisms along the line of sight can cause fluctuations in the \HI\ distribution, which to first order can be described as
$n_{\mathrm{H\,I}}(z,\hat{n})=\bar{n}_{\mathrm{H\,I}}(z)\left[1+\delta_{\mathrm{H\,I}}(z,\hat{n})\right]$
in direction $\hat{n}$. Consequently, we have a first-order temperature perturbation,
$T_{\mathrm{H\,I}}(z,\hat{n})=\bar{T}_{\mathrm{H\,I}}(z)\left[1+\delta T_{\mathrm{H\,I}}(z,\hat{n})\right]$.
Recovering this perturbation term is one of the main BINGO goals, as discussed in \citep{BINGO_VII}. We can describe the first-order perturbation in the conformal Newtonian gauge as a sum of different physical effects:
\begin{align}
\delta T_{\mathrm{H\,I}}(z,\hat{n})
&=
\delta_{\mathrm{n}}
-\frac{1}{\mathcal{H}}\,\hat{\mathbf{n}}\cdot\left(\hat{\mathbf{n}}\cdot \nabla\hat{\mathbf{v}}\right)
\nonumber\\
&\quad+
\left(\frac{\mathrm{d}}{\mathrm{d}\eta}\ln\left(a^3\bar{n}_{\mathrm{H\,I}}\right)
-\frac{\dot{\mathcal{H}}}{\mathcal{H}}-2\mathcal{H}\right)\delta \eta
+\frac{1}{\mathcal{H}}\dot{\Phi}+\Psi
\,,
\label{eqn: deltaT}
\end{align}
where the first term expresses the contribution from the perturbation of the \HI\ density, and the second term is related to redshift-space distortions (RSD) arising from peculiar velocities. The third term (in parentheses) accounts for evaluating the background brightness temperature at the perturbed conformal time $\eta$ of the observed redshift. The last two terms arise in the Newtonian gauge, where $\Psi$ is the gravitational potential and $\Phi$ is the perturbation of the metric potential. The penultimate term in Eq.~\ref{eqn: deltaT} is associated with the integrated Sachs--Wolfe (ISW) effect, and the final term relates to the conversion between redshift increments and radial distances in the gas frame. Here $\mathcal{H}$ is the Hubble function in conformal time (see also \citep{hall2013}   ).

\subsection{Foreground signals}
The 21\,cm signal is extremely weak compared to other astrophysical and cosmological signals detected by the receivers. BINGO operates at much lower frequencies than \emph{Planck} \citep{planck}. In this band, the main sources of sky emission in addition to the 21\,cm signal are commonly referred to as \emph{foregrounds}. They can be classified into Galactic, extragalactic, and cosmological emissions.

Galactic emissions generally originate from the Galaxy's interstellar medium (ISM), constituted by cold molecular and atomic clouds. The medium between these clouds is partially ionized, likely by supernovae. These environments are heavily concentrated in the Galactic plane; consequently, as seen in Fig.~\ref{fig: Foregrounds}, significant radio emission originates from this region. The strongest emissions in this frequency range are synchrotron radiation, produced when energetic charged particles move through the Galactic magnetic field; free--free emission, produced when free electrons interact with ions in the ionized medium; and anomalous microwave emission (AME), likely due to spinning dust grains.

Extragalactic radio emission is mainly produced by active galactic nuclei (AGN) and star-forming galaxies (SFG) \citep{chapman2019foregrounds}. AGN emit synchrotron radiation via matter accretion onto central supermassive black holes, ejecting jets perpendicular to the accretion plane. SFGs produce synchrotron emission (similar to our Galaxy) and free--free emission from regions of ionized hydrogen. In this work, we model extragalactic emission as radio point sources. This component comprises two types of unresolved sources: a population of faint objects that cannot be resolved individually, and the brightest extragalactic objects identified from catalogs at different frequencies. Finally, we assume the cosmological signal contribution is dominated by the cosmic microwave background (CMB).

    \section{Foreground Removal filter}
\label{Section: AppendixB}

For estimating the diffuse foreground emissions, we need to infer their spectral evolution matrix $\mathbf{A}$ (as well as $\mathbf{S}$) through a linear inversion problem. If the number of channels equals the number of sources ($N_{\mathrm{ch}}=n_{\mathrm{s}}$) and noise is absent, the mixing matrix is square and invertible, providing a unique solution $\mathbf{A}^{-1}$. However, due to unmodeled signals and the fact that the mixing-matrix dimension is rarely equal to the number of channels, $\mathbf{A}$ is generally not invertible. The method used to estimate this matrix characterizes each algorithm. A general approach is to find a filter $\mathbf{W}$ such that, when applied to the observation matrix $\mathbf{X}$, it yields an estimate of the source (foreground-template) matrix. This filter effectively downweights $\mathbf{N}$ while preserving the foreground signals:
\begin{align}
\hat{\mathbf{S}} &= \mathbf{W}\mathbf{X} \nonumber\\
&= (\mathbf{W}\mathbf{A})\mathbf{S} + \mathbf{W}\mathbf{N}.
\end{align}

Following \citep{hyvarinen2001}, a natural approach is to minimize the contribution of $\mathbf{N}$ using a least-squares criterion. This deterministic approach requires no assumptions about the probability distribution:
\begin{equation}
\nabla_{\mathbf{S}}\bigg|_{\hat{\mathbf{A}},\hat{\mathbf{S}}}
\left( \frac{1}{2}\,\|\mathbf{N}\|^2 \right)= 0,
\end{equation}
which yields the \emph{Moore--Penrose pseudoinverse}, $\mathbf{A}^{+}$:
\begin{equation}
\mathbf{W} = \mathbf{A}^+ = (\hat{\mathbf{A}}^{\mathsf{T}}\hat{\mathbf{A}})^{-1}\hat{\mathbf{A}}^{\mathsf{T}}.
\label{eqn:W_LS}
\end{equation}

We used the canonical metric with the $\ell_2$ norm, $\|\cdot\|=\|\cdot\|_{\ell_2}$.\footnote{For an arbitrary matrix $\mathbf{B}$, the canonical metric is defined as $\|\mathbf{B}\|^2=\mathbf{B}^{\mathsf{T}}\mathbf{B}$.} More generally, one can employ a weighted metric, e.g.\ $\|\cdot\|_{\mathbf{Q}}$,\footnote{For arbitrary matrices $\mathbf{B}$ and $\mathbf{Q}$, $\|\mathbf{B}\|_{\mathbf{Q}}^2=\mathbf{B}^{\mathsf{T}}\mathbf{Q}\mathbf{B}$.} where $\mathbf{Q}$ is a weighting matrix. In this case, the minimization generalizes to \emph{generalized least squares} (GLS):
\begin{equation}
\nabla_{\mathbf{S}}\bigg|_{\hat{\mathbf{A}},\hat{\mathbf{S}}}
\left( \frac{1}{2}\,\|\mathbf{N}\|_{\mathbf{Q}}^2 \right)= 0.
\end{equation}
The resulting filter is
\begin{equation}
\mathbf{W} = (\hat{\mathbf{A}}^{\mathsf{T}}\mathbf{Q}\hat{\mathbf{A}})^{-1}\left(\hat{\mathbf{A}}^{\mathsf{T}}\mathbf{Q}\right),
\label{eqn:W_GLS}
\end{equation}
which is also known as the \emph{Gauss--Markov estimator} in information theory.

We define the residual matrix $\mathbf{R}$ as the difference between the observation matrix and the estimated foreground matrix,
\begin{equation}
\mathbf{R} = \mathbf{X} - \hat{\mathbf{A}}\hat{\mathbf{S}}
= (\mathbf{I}-\mathbf{W}_{\mathrm{FG}})\mathbf{X},
\end{equation}
where $\mathbf{W}_{\mathrm{FG}} \doteq \hat{\mathbf{A}}\mathbf{W}$. The components of $\mathbf{W}_{\mathrm{FG}}$ typically determine how each frequency (channel) contributes to the estimation of the neutral-hydrogen spectrum \citep{marins2024}.

    \section{Algorithms: FastICA, GMCA, and GNILC}
\label{Section: AppendixC}

\subsection{Fast Independent Component Analysis}
\emph{Independent Component Analysis} (ICA) is an algorithm applied to astrophysical (and cosmological) observations to model or remove foregrounds using the hypothesis that astrophysical sources are statistically (and mutually) independent. Signals from different sources are statistically independent, meaning they do not contain information about one another. Mathematically, statistical independence implies that the joint probability distribution factorizes into the product of the marginals.

ICA-based algorithms look for a linearly transformed matrix $\hat{\mathbf{S}}=\mathbf{W}\mathbf{X}$, where each row is a transformed vector and all components are mutually independent \citep{hyvarinen1999}. In this work we use an ICA-based approach that does not explicitly model instrumental noise: ICA is used to model the foregrounds, while non-smooth spectral components are expected to remain in the residuals. Let $\mathbf{w}_{i}$ be the $i$th row of $\mathbf{W}$, i.e.
$\mathbf{W} = \left[\mathbf{w}_1,\mathbf{w}_2,\dots,\mathbf{w}_{n_{\mathrm{s}}}\right]^{\mathrm{T}}$.
We can write an independent component as
\begin{equation}
\mathbf{y}_i=\sum_{j=1}^{N_{\mathrm{ch}}}W_{ij}\mathbf{x}_j=\mathbf{w}_{i}^{\mathrm{T}}\mathbf{X},
\end{equation}
which is equivalent to searching for a maximally non-Gaussian component \citep{hyvarinen2001}. At each iteration, the algorithm searches for a new transformed variable that is more independent than the previous one, in the sense of maximizing a non-Gaussianity criterion.

If we consider $\mathbf{y}_i=\mathbf{w}_{i}^{\mathrm{T}}\mathbf{X}=(\mathbf{w}_{i}^{\mathrm{T}}\mathbf{A})\mathbf{s}$, then for an idealized noise-free and invertible case one would have $(\mathbf{w}_{i}^{\mathrm{T}}\mathbf{A})^{\mathrm{T}}$ selecting a single component of $\mathbf{s}$. In practice, however, the mixing matrix may not be invertible and the data are not noise-free, so $\mathbf{w}_{i}$ must be estimated. By the central limit theorem, a linear combination of independent components tends to be more Gaussian than the components themselves; therefore, the least Gaussian projection is (approximately) achieved when the projection is close to one of the original components. This motivates estimating $\mathbf{w}_i$ by maximizing non-Gaussianity.

\citep{hyvarinen2001} proposed \emph{negentropy} as a robust measure of non-Gaussianity (often preferable to kurtosis). Negentropy is defined as
\begin{equation}
J(\boldsymbol{\xi}) = H(\boldsymbol{\xi}_{\mathrm{G}}) - H(\boldsymbol{\xi}),
\end{equation}
where $H$ is the entropy, $\boldsymbol{\xi}$ is a random vector, and the subscript $\mathrm{G}$ denotes a Gaussian random variable with the same covariance matrix as $\boldsymbol{\xi}$. Since negentropy is computationally expensive to evaluate exactly, it is typically approximated using a \emph{contrast function} $g$:
\begin{equation}
J(\boldsymbol{\xi}) \propto
\Big[\mathbb{E}\{g(\boldsymbol{\xi})\}-\mathbb{E}\{g(\nu_{\mathrm{G}})\}\Big]^2,
\end{equation}
for a non-quadratic function $g$, and a Gaussian variable $\nu_{\mathrm{G}}$ with zero mean and unit variance. In this work we use FastICA, a fixed-point algorithm that maximizes an approximation to negentropy, with $\xi=\mathbf{w}^{\mathrm{T}}\mathbf{z}$ where $\mathbf{z}$ is the whitened version of $\mathbf{x}$.

\subsubsection{FastICA as an optimization problem}
FastICA can be described as an optimization problem: it maximizes non-Gaussianity (via negentropy) under an orthogonality constraint on $\mathbf{W}$. We write
\begin{equation}
\{\mathbf{W}\}
=
\underset{\mathbf{W}}{\mathrm{argmax}}
\left\{
J\!\left(\mathbf{W}^{\mathrm{T}}\mathbf{X}\right)
+\lambda\,\big\|\mathbf{W}\mathbf{W}^{\mathrm{T}}-\mathbf{I}\big\|
\right\},
\end{equation}
where $\mathrm{argmax}$ returns the matrix $\mathbf{W}$ that maximizes the expression inside braces.

FastICA was introduced for the separation of astrophysical components in \citep{maino2002}. It has been used in 21\,cm cosmology in both reionization \citep{chapman2012} and post-reionization \citep{carucci2020} contexts. FastICA estimates neither the 21\,cm signal nor thermal noise explicitly: these remain in the residuals after foreground estimation. FastICA yields solutions that are unique up to the standard ICA ambiguities (permutation and scaling).

\subsection{Generalized Morphological Component Analysis}
Generalized Morphological Component Analysis (GMCA) is based on two ideas: \emph{morphological diversity} and \emph{sparsity}. We briefly describe these properties below.

\subsubsection{Sparsity and morphological diversity}
A signal $\mathbf{y}$ is said to be \emph{sparse} in a dictionary\footnote{A dictionary $\mathcal{D}$ is a collection of parameterized waveforms $\{\phi_{\gamma};\,\gamma\in\Gamma\}$. A signal $s$ can be approximated as $s=\sum_{\gamma\in\Gamma}\alpha_{\gamma}\phi_{\gamma}+\mathrm{Res}$, where $\mathrm{Res}$ denotes the residual. Examples of dictionaries include Fourier, wavelet, and Gabor bases. See, e.g., \citep{chen1995,chen2001}.} $\boldsymbol{\Phi}$ if it can be represented by only a few elements of $\boldsymbol{\Phi}$. For instance,
\begin{equation}
\mathbf{y}= \boldsymbol{\alpha}\boldsymbol{\Phi} = \sum_{\gamma \in \Gamma}\alpha^{\gamma}\phi_{\gamma},
\end{equation}
where $\alpha^{\gamma}$ is the coefficient associated with waveform $\phi_{\gamma}$. If $\mathbf{y}$ is sparse in $\boldsymbol{\Phi}$, the number of nonzero coefficients is small. More precisely, $\mathbf{y}$ is $k$-sparse in $\boldsymbol{\Phi}$ if the number of non-vanishing coefficients is $k$. In practice, signals are rarely strictly sparse, so it is important to find representations in which they are \emph{approximately} sparse. Wavelets are useful for this purpose because they are localized in both space (or time) and frequency domains \citep{starckWT1997,starckIUWT2007}. Wavelets on the sphere are also localized in both harmonic and pixel domains and are widely used to analyze astrophysical maps.

The use of sparsity for blind source separation (BSS) was introduced in \citep{Zibulevsky2000}. The morphological component analysis (MCA) framework is described in \citep{starckMCA2004redundant}. Its multichannel extension leads to multichannel MCA (MMCA), which assumes that each source signal $s_j$ can be sparsely represented in a dictionary $\boldsymbol{\Phi}_j$. Finally, \citep{bobin2007} introduced GMCA, extending MMCA by allowing each $s_j$ to be represented in a \emph{superdictionary} $\mathcal{D}=[\boldsymbol{\Phi}_1^{\mathrm{T}},\dots,\boldsymbol{\Phi}_D^{\mathrm{T}}]^{\mathrm{T}}$. Each source $s_j$ has a representation
\begin{align}
s_j &= \sum_{k=1}^{D}\phi_{jk} = \sum_{k=1}^{D}\alpha_{j}^{\ k}\boldsymbol{\Phi}_{k},
\end{align}
or, equivalently, in matrix form,
\begin{equation}
\mathbf{S}= \boldsymbol{\alpha}\mathcal{D}.
\end{equation}
Thus $s_j$ is modeled as a linear combination of $D$ morphological components $\{\phi_{jk}\}_{k=1}^{D}$, each sparse in a given orthogonal basis.

\subsubsection{GMCA as an optimization problem}
Assuming that each row of $\mathbf{S}$ is sparse in $\mathcal{D}$ leads to non-uniqueness because $\mathcal{D}$ is overcomplete. To select a solution, one typically searches for the \emph{sparsest} representation. The original formulation uses an $\ell_0$ penalty, but this is non-convex and computationally expensive. A common relaxation replaces $\|\alpha\|_{\ell_0}$ with an $\ell_1$ penalty, yielding
\begin{equation}
\{\hat{\mathbf{A}},\hat{\boldsymbol{\alpha}}\}
=
\underset{\mathbf{A},\boldsymbol{\alpha}}{\mathrm{argmin}}
\left\{
\|\mathbf{X}-\mathbf{A}\boldsymbol{\alpha}\mathcal{D}\|^2_{\mathrm{F},\mathbf{C}_{\mathbf{N}}^{-1}}
+2\lambda\sum_{j=1}^{n_{\mathrm{s}}}\|\alpha_{j}\|_{\ell_1}
\right\},
\end{equation}
where the first term is the $\mathbf{C}_{\mathbf{N}}^{-1}$-weighted Frobenius norm\footnote{Let $\mathbf{Z}$ be a matrix. The $\mathbf{Q}$-weighted Frobenius norm is $\|\mathbf{Z}\|_{\mathrm{F},\mathbf{Q}}^2=\mathrm{Tr}\!\left(\mathbf{Z}\mathbf{Q}^{-1}\mathbf{Z}^{\mathrm{T}}\right)$. If $\mathbf{Q}$ is the identity, this reduces to $\|\mathbf{Z}\|_{\mathrm{F}}^2=\mathrm{Tr}(\mathbf{Z}\mathbf{Z}^{\mathrm{T}})$.} and $\lambda$ is a regularization parameter that controls the thresholding. We do not detail the full algorithm here; see \citep{bobin2007,GMCA2008SZ}. In brief, at iteration $i$ one estimates $\mathbf{S}^{(i)}$ given $\mathbf{A}^{(i-1)}$, and then updates $\mathbf{A}^{(i)}$ given $\mathbf{S}^{(i)}$, typically using thresholding in the sparse domain.

\subsection{Generalized Needlet Internal Linear Combination}
\label{sec:GNILC0}
Since the effective ratio between the 21\,cm signal and the total observed signal depends on sky direction, the number of (non-physical) templates required to describe the foreground contribution varies across the sky. This can be seen by comparing directions close to the Galaxy to those outside it. Although thermal noise and the 21\,cm signal are close to Gaussian across channels, foregrounds are not, and the relative contributions depend on frequency. A localized description of the foreground contribution in both spatial and spectral domains is therefore essential. By moving part of the analysis to harmonic space, we can exploit localized filtering with needlets \citep{NeedletsCMB2008}.

Needlet-ILC (NILC) was introduced to extract CMB maps from WMAP data in needlet space \citep{Delabrouille2009}. \citep{Remazeilles:2011} generalized this approach to a multidimensional filter, leading to the GNILC algorithm, which models the observation as a set of multidimensional components (templates) rather than a single template.

\subsubsection{GNILC as an optimization problem}
\label{sec:GNILC}
Roughly speaking, GNILC can be viewed as the search for a filter with unit response to the target subspace while suppressing other components. Following \citep{Remazeilles:2011}, it can be written as an optimization problem:
\begin{equation}
\{\hat{\mathbf{A}},\hat{\mathbf{S}}\}
=
\underset{\mathbf{A},\mathbf{S}}{\mathrm{argmin}}
\left\{
\|\mathbf{W}\|^2_{\mathrm{F},\mathbf{R}_{\mathbf{X}}^{-1}}
+\big\|\boldsymbol{\Lambda}^{\mathrm{T}}(\mathbf{A}-\mathbf{W}\mathbf{A})\big\|_{\mathrm{F}}
\right\},
\end{equation}
where $\boldsymbol{\Lambda}$ is a matrix of Lagrange multipliers. The solution corresponds to Eq.~\ref{eqn:W_GLS} with $\mathbf{Q}=\mathbf{R}_{\mathbf{X}}^{-1}$. GNILC implements this in needlet space by using covariance matrices of needlet coefficients in bands of multipoles, estimating a filter per needlet band, and then transforming back to pixel space.

GNILC uses a prior angular power spectrum of the target signal to estimate the signal-to-noise ratio and to separate the degrees of freedom of the target and unwanted signals. In other words, it uses a prior template of the 21\,cm signal to estimate the foreground degrees of freedom $N_{\mathrm{s}}$ in the observation, per needlet band and pixel.

GNILC was first applied to 21\,cm maps in \citep{olivari2016}. In that work, $N_{\mathrm{s}}$ denotes the degrees of freedom of foregrounds plus instrumental noise and is estimated using the Akaike information criterion (AIC) \citep{AIC1974,AICbozdogan1987model}. Here we separate the foregrounds from the 21\,cm signal and (instrumental) noise, projecting them onto a $(N_{\mathrm{ch}}-N_{\mathrm{s}})$-dimensional subspace. GNILC performs a singular value decomposition (SVD) of the whitened data covariance matrix,
\begin{equation}
\hat{\mathbf{R}}_{\mathrm{prior}}^{-1/2}\,\mathbf{R}_{\mathbf{X}}\,\hat{\mathbf{R}}_{\mathrm{prior}}^{-1/2}
=
\hat{\mathbf{U}}\,\hat{\mathbf{D}}\,\hat{\mathbf{U}}^{\dagger},
\label{eq: SVDdec}
\end{equation}
where $\hat{\mathbf{R}}_{\mathrm{prior}}$ is the covariance matrix of the prior template, and $\hat{\mathbf{U}}$ and $\hat{\mathbf{D}}$ contain eigenvectors and eigenvalues of the whitened covariance, respectively. The eigenvector matrix is split into foreground and target subspaces,
$\hat{\mathbf{U}}=\left[\hat{\mathbf{U}}_{\mathrm{FG}},\,\hat{\mathbf{U}}_{\mathrm{H\,I}+\mathrm{N}}\right]$.
The AIC determines the dimension of the $(\mathrm{H\,I}+\mathrm{N})$ subspace, i.e.\ the number of columns of $\hat{\mathbf{U}}_{\mathrm{H\,I}+\mathrm{N}}$. GNILC then obtains an estimate of the mixing matrix as
\begin{equation}
\hat{\mathbf{A}}
=
\hat{\mathbf{R}}_{\mathrm{prior}}^{1/2}\,\hat{\mathbf{U}}_{\mathrm{H\,I}+\mathrm{N}}.
\end{equation}

With the mixing matrix estimated, GNILC computes a multidimensional ILC filter per pixel and needlet band (indexed by $j$),
\begin{equation}
\hat{\mathbf{x}}^{(j)}_{\mathrm{H\,I}+\mathrm{N}}(p) = \mathbf{W}^{(j)}(p)\,\mathbf{y}^{(j)}(p),
\end{equation}
where $\hat{\mathbf{x}}^{(j)}_{\mathrm{H\,I}+\mathrm{N}}(p)$ is the multichannel $(\mathrm{H\,I}+\mathrm{N})$ estimate at pixel $p$ in band $j$. After processing all needlet bands, GNILC performs the inverse needlet transform (INT) to obtain the $(\mathrm{H\,I}+\mathrm{N})$ maps in pixel space,
\begin{equation}
\{\hat{\mathbf{x}}^{(j)}_{\mathrm{H\,I}+\mathrm{N}}(p)\}_j
\stackrel{\mathrm{INT}}{\longrightarrow}
\hat{\mathbf{x}}_{\mathrm{H\,I}+\mathrm{N}}(p).
\end{equation}

For a complete description of the algorithm, see \citep{BINGO_V}. As a prior, we used a template for each channel from an arbitrary realization of 21\,cm plus white noise.

\begin{figure}
\centering
    \includegraphics[width=\textwidth]{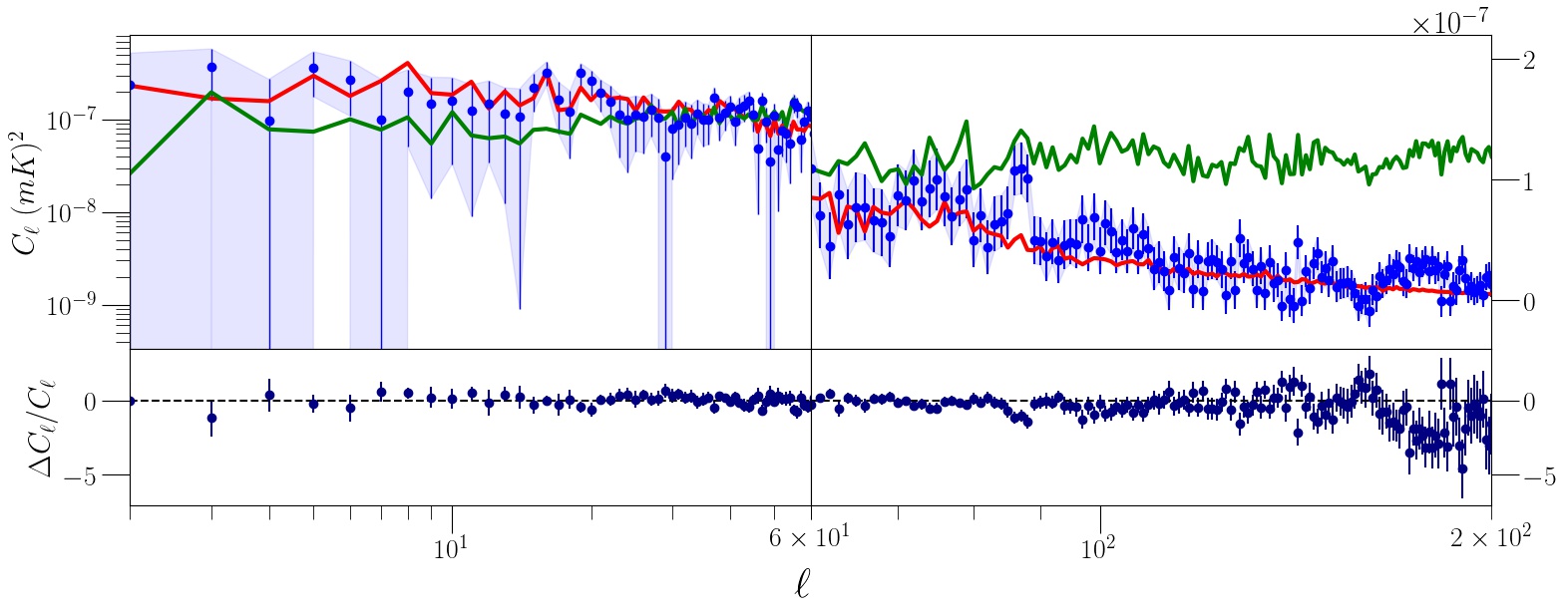}\\
    \includegraphics[width=\textwidth]{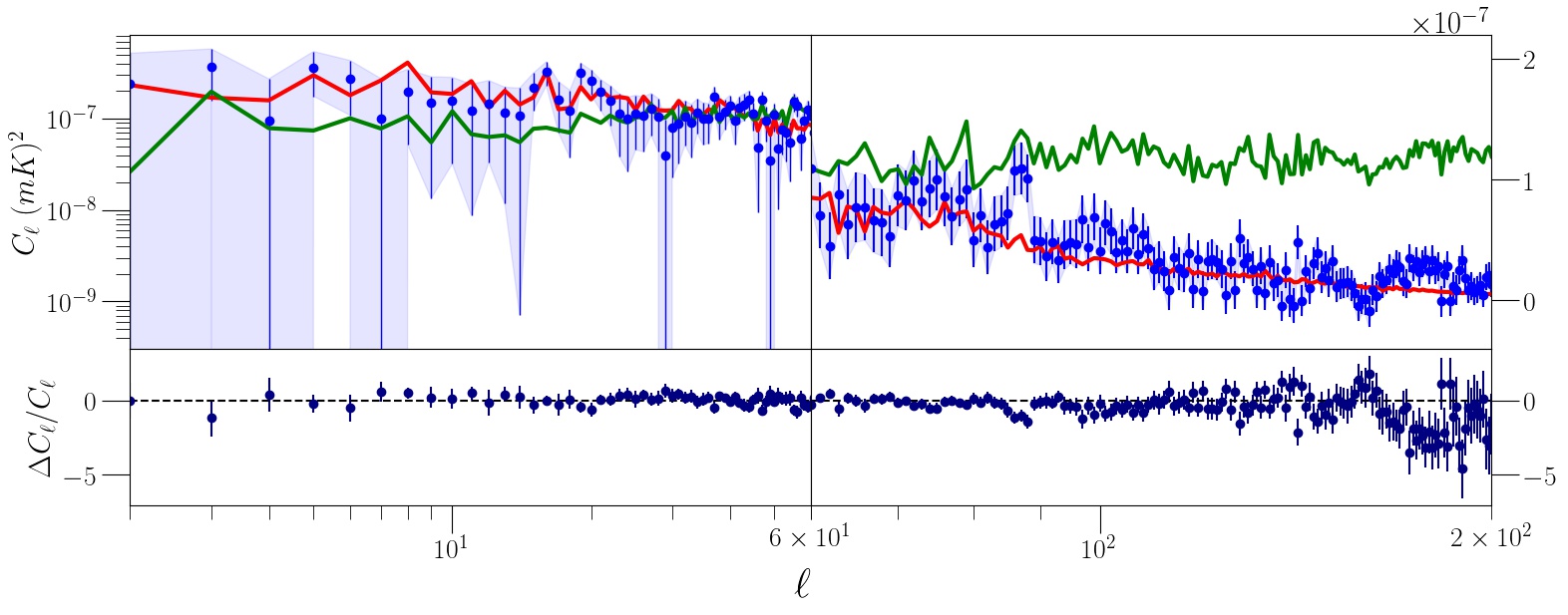}\\
    \includegraphics[width=\textwidth]{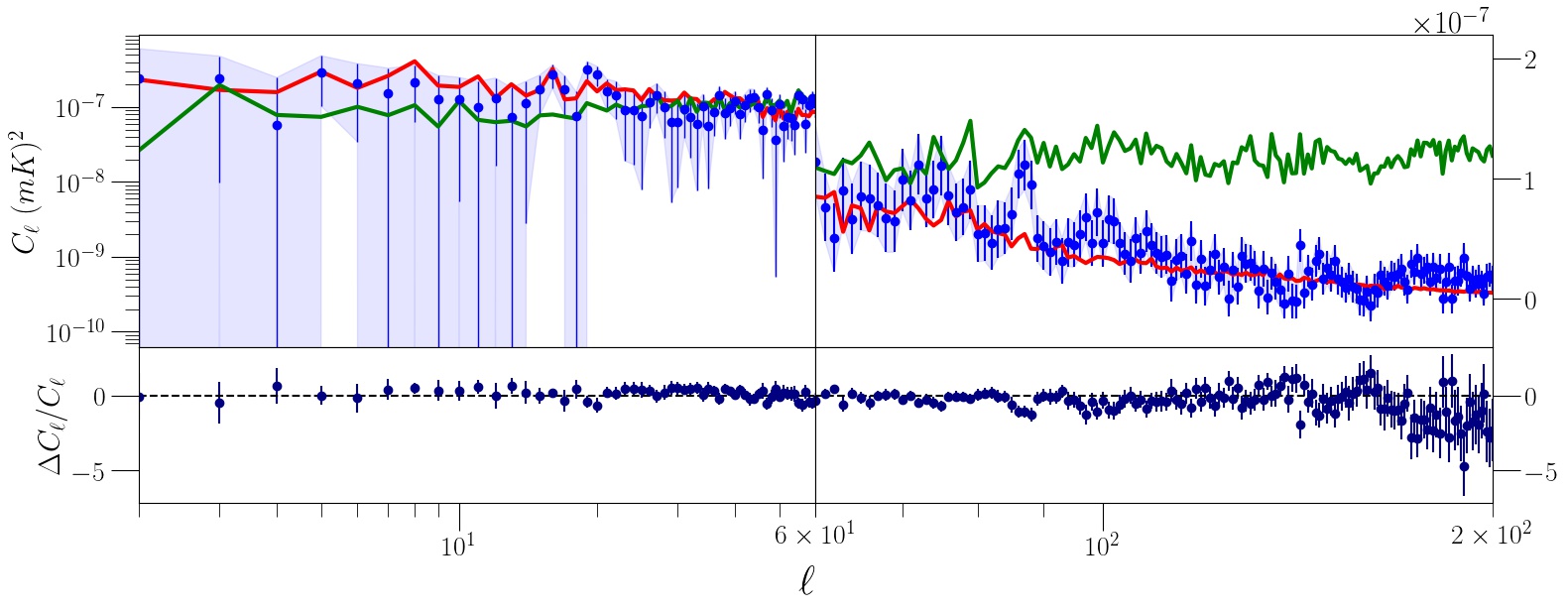}
\caption{Angular power spectra for the 10th channel using FastICA (top), GMCA (middle), and GNILC (bottom). The analysis uses 400 realizations, with $n_{\mathrm{s}}=3$ for FastICA and GNILC. The curves show the estimated \HI\ (blue), the input \HI\ (red), and white noise (green). Below each graph is the residual difference between the estimated and input \HI\ spectra.}
\label{fig: Algorithms_HI_APS_estimation}
\end{figure}

    \bibliographystyle{JHEP}
    \bibliography{biblio}
    
    \end{document}